\documentclass[aps,pra,preprint]{revtex4-1}

\usepackage{dcolumn}
\usepackage{bm}

\usepackage{graphicx}

\usepackage{color}
\definecolor{zema}{rgb}{0,0,0}
\definecolor{zema2}{rgb}{0,0,0}

\usepackage{amsmath,amsthm,amssymb,amsfonts}
\usepackage{units}

\usepackage{lineno}

\newcommand{\mrm}{\mathrm}

\begin{document}


\newlength{\OneColumnPictureWidth}
\makeatletter%
\setlength{\OneColumnPictureWidth}{\twocolumn@sw{\columnwidth}{0.48\textwidth}}
\makeatother%

\title{Dynamic stabilization of the magnetic field surrounding the neutron 
electric dipole moment spectrometer at the Paul Scherrer Institute}

\author{S.~Afach}
\affiliation{Paul Scherrer Institute, Villigen, Switzerland}
\affiliation{Institute for Particle Physics, Eidgen\"ossische Technische Hochschule, Z\"urich,  Switzerland}

\author{G.~Bison}
\affiliation{Paul Scherrer Institute, Villigen, Switzerland}

\author{K.~Bodek}
\affiliation{Jagellonian University, Cracow, Poland}

\author{F.~Burri}
\affiliation{Paul Scherrer Institute, Villigen, Switzerland}

\author{Z.~Chowdhuri}
\affiliation{Paul Scherrer Institute, Villigen, Switzerland}

\author{M.~Daum}
\affiliation{Paul Scherrer Institute, Villigen, Switzerland}

\author{M.~Fertl}
\altaffiliation{present address: University of Washington, Seattle, WA, USA}
\affiliation{Paul Scherrer Institute, Villigen, Switzerland}
\affiliation{Institute for Particle Physics, Eidgen\"ossische Technische Hochschule, Z\"urich,  Switzerland}

\author{B.~Franke}
\email[corresponding author: ]{beatrice.franke@psi.ch}
\altaffiliation{present address: Max Planck Institute of Quantum Optics, Garching, Germany}
\affiliation{Paul Scherrer Institute, Villigen, Switzerland}
\affiliation{Institute for Particle Physics, Eidgen\"ossische Technische Hochschule, Z\"urich,  Switzerland}

\author{Z.~Grujic}
\affiliation{University of Fribourg, Switzerland}

\author{V.~H\'{e}laine}
\affiliation{Laboratoire de Physique Corpusculaire, Caen, France}
\affiliation{Paul Scherrer Institute, Villigen, Switzerland}

\author{R.~Henneck}
\affiliation{Paul Scherrer Institute, Villigen, Switzerland}

\author{M.~Kasprzak}
\affiliation{University of Fribourg, Switzerland}

\author{K.~Kirch}
\affiliation{Paul Scherrer Institute, Villigen, Switzerland}
\affiliation{Institute for Particle Physics, Eidgen\"ossische Technische Hochschule, Z\"urich,  Switzerland}

\author{H.-C.~Koch}
\affiliation{Institut f\"ur Physik, Johannes-Gutenberg-Universit\"at, Mainz, Germany}
\affiliation{University of Fribourg, Switzerland}

\author{A.~Kozela}
\affiliation{Henryk Niedwodnicza\'nski Institute for Nuclear Physics, Cracow, Poland}

\author{J.~Krempel}
\affiliation{Institute for Particle Physics, Eidgen\"ossische Technische Hochschule, Z\"urich,  Switzerland}

\author{B.~Lauss}
\email[corresponding author: ]{bernhard.lauss@psi.ch}
\affiliation{Paul Scherrer Institute, Villigen, Switzerland}

\author{T.~Lefort}
\affiliation{Laboratoire de Physique Corpusculaire, Caen, France}

\author{Y.~Lemi\`ere}
\affiliation{Laboratoire de Physique Corpusculaire, Caen, France}

\author{M.~Meier}
\affiliation{Paul Scherrer Institute, Villigen, Switzerland}

\author{O.~Naviliat-Cuncic}
\altaffiliation{present address: Michigan State University, East Lansing, MI, USA}
\affiliation{Laboratoire de Physique Corpusculaire, Caen, France}

\author{F.M.~Piegsa}
\affiliation{Institute for Particle Physics, Eidgen\"ossische Technische Hochschule, Z\"urich,  Switzerland}

\author{G.~Pignol}
\affiliation{Laboratoire de Physique Subatomique et de Cosmologie, Grenoble, France}

\author{C.~Plonka-Spehr}
\affiliation{Institut f\"ur Kernchemie, Johannes-Gutenberg-Universit\"at, Mainz, Germany}

\author{P.~N.~Prashanth}
\affiliation{Katholieke Universiteit, Leuven, Belgium}
\affiliation{Paul Scherrer Institute, Villigen, Switzerland}

\author{G.~Qu\'em\'ener}
\affiliation{Laboratoire de Physique Corpusculaire, Caen, France}

\author{D.~Rebreyend}
\affiliation{Laboratoire de Physique Subatomique et de Cosmologie, Grenoble, France}

\author{S.~Roccia}
\affiliation{Centre de Sciences Nucl\'eaires et de Sciences de la Mati\`ere, 
Universit\'e Paris Sud-CNRS/IN2P3, Orsay, France}

\author{P.~Schmidt-Wellenburg}
\affiliation{Paul Scherrer Institute, Villigen, Switzerland}

\author{A.~Schnabel}
\affiliation{Physikalisch Technische Bundesanstalt, Berlin, Germany}

\author{N.~Severijns}
\affiliation{Katholieke Universiteit, Leuven, Belgium}

\author{J.~Voigt}
\affiliation{Physikalisch Technische Bundesanstalt, Berlin, Germany}

\author{A.~Weis}
\affiliation{University of Fribourg, Switzerland}

\author{G.~Wyszynski}
\affiliation{Jagellonian University, Cracow, Poland}
\affiliation{Institute for Particle Physics, Eidgen\"ossische Technische Hochschule, Z\"urich,  Switzerland}

\author{J.~Zejma}
\affiliation{Jagellonian University, Cracow, Poland}

\author{J.~Zenner}
\affiliation{Institute for Particle Physics, Eidgen\"ossische Technische Hochschule, Z\"urich,  Switzerland}

\author{G.~Zsigmond}
\affiliation{Paul Scherrer Institute, Villigen, Switzerland}

\begin{abstract}
The Surrounding Field Compensation (SFC) system described in this work is installed around 
the four-layer Mu-metal magnetic shield of the neutron electric dipole moment spectrometer 
located at the Paul Scherrer Institute. 
The SFC system reduces the DC component of the external magnetic field by a factor of about 20. 
Within a control volume 
of approximately  2.5\,m\,$\times$\,2.5\,m\,$\times$\,3\,m
disturbances of the magnetic field are attenuated by 
factors of 5 to 50 at a bandwidth from $10^{-3}$\,Hz
up to 0.5\,Hz,
which corresponds to integration times longer 
than several hundreds of seconds and represent the important timescale for the nEDM measurement.
These shielding factors apply to random environmental noise from arbitrary sources.
This is achieved via a proportional-integral feedback stabilization system that includes a regularized 
pseudoinverse matrix of proportionality factors which correlates magnetic field changes at 
all sensor positions to current changes in the SFC coils.
\end{abstract}



\date{\today}

\maketitle


\section{Introduction}
\label{sec:intro}

In this work we describe the setup and performance of a stabilization system 
which monitors
the environmental magnetic field 
and compensates 
for 
magnetic disturbances 
at several points around the respective control volume of roughly 10\,m$^3$ in size.
This surrounding field compensation (SFC) system 
is an
important part of 
the neutron electric dipole moment (nEDM) 
experiment \cite{PPnedmPSI,Baker2013}, 
located at the ultracold neutron (UCN) source \cite{psiucnsource,Lauss2014}
of the Paul Scherrer Institute (PSI) 
in Villigen, Switzerland.

Ultracold neutrons have very low energies, below about 300\,neV, 
and thus can be trapped in bottles and observed for times comparable to the lifetime of the free neutron. 
This fact makes them an excellent tool to search for a possible electric dipole moment of the neutron
~\cite{nEDM-Baker2006,serebrov2014,PPnedmPSI}, 
which is considered to be one of the 
most important experiments in particle physics 
(see e.g.~\cite{nedm-review,Raidal2008,strategy2013})
and will contribute to 
answering
the fundamental questions on the origin of the
matter-antimatter asymmetry observed in our universe.
An introduction to UCN and the nEDM can be found in \cite{golub}.

The nEDM experiment at PSI applies the 
Ramsey method of separated oscillatory fields \cite{PhysRev.76.996} 
to spin-polarized UCN confined in a precession chamber 
located in the center of a vacuum tank inside a four layer Mu-metal magnetic shield. 
The sensitivity of the experiment depends on the stability of the internal magnetic field 
and field gradients inside the UCN storage chamber.
Thus, of paramount importance for the measurement sensitivity are 
(i) a four-layer cylindrical magnetic shield made 
of Mu-metal (a high permeability NiFe-alloy) around the vacuum tank, and
(ii) very high -- tens of femtotesla -- precision magnetometers in and 
around the UCN storage chamber
to monitor the internal magnetic field changes.
The stability and homogeneity of the magnetic field inside the shield 
depends to a large extent on the magnetization state of the Mu-metal.
In order to maintain a stable magnetization,
the surrounding magnetic field must be as stable as possible for 
magnetic field changes with frequencies
below a few Hz.
To fullfill this task the SFC system was set up around the Mu-metal shield. 
It provides a static compensation of the Earth's magnetic field
and additionally
a dynamic compensation for the environmental magnetic field changes. 

Our distinctive approach to the SFC system, 
described in detail in \cite{DissBea}, 
uses a regularized pseudoinverse 
matrix of proportionality factors in the feedback algorithm
and thereby
avoids 
introducing noise in orthogonal directions (as e.g.\,in \cite{Belfi2010}).
This allows us to stabilize the 
magnetic field simultaneously at many positions
within the control volume.
Our approach builds on 
previous experimental efforts at PSI \cite{PSIMagstab2005}
which used a simple dynamic magnetic field stabilization system. 
An overview of
other magnetic field compensation systems published before the year 2005 
can be found in \cite{PSIMagstab2005}. 

While
active surrounding magnetic field compensation is a necessity for 
high-sensitivity nEDM searches, 
it is also used in other particle physics experiments 
in order to achieve isotropic detector performance \cite{Bodmer2013}, 
in bio-magnetism \cite{Baltag2010,Baltag2012}
and in medical research \cite{Naka2009, Dong2013}
to improve signal and image quality.

This article is arranged in the following sections: 
Sec.\,\ref{sec:char}, 
characteristics of the SFC system;
Sec.\,\ref{sec:Inplusone}, feedback algorithm for dynamic compensation;
Sec.\,\ref{sec:shieldingfactor},
method used for quantifying the performance via a shielding factor; 
Sec.\,\ref{sec:results},
the system performance,
and 
Sec.\,\ref{sec:conclusion},
conclusions and outlook.


\section{System characteristics}
\label{sec:char}

\subsection{Overview}

\begin{figure}
\centering
   \includegraphics[width=\OneColumnPictureWidth]{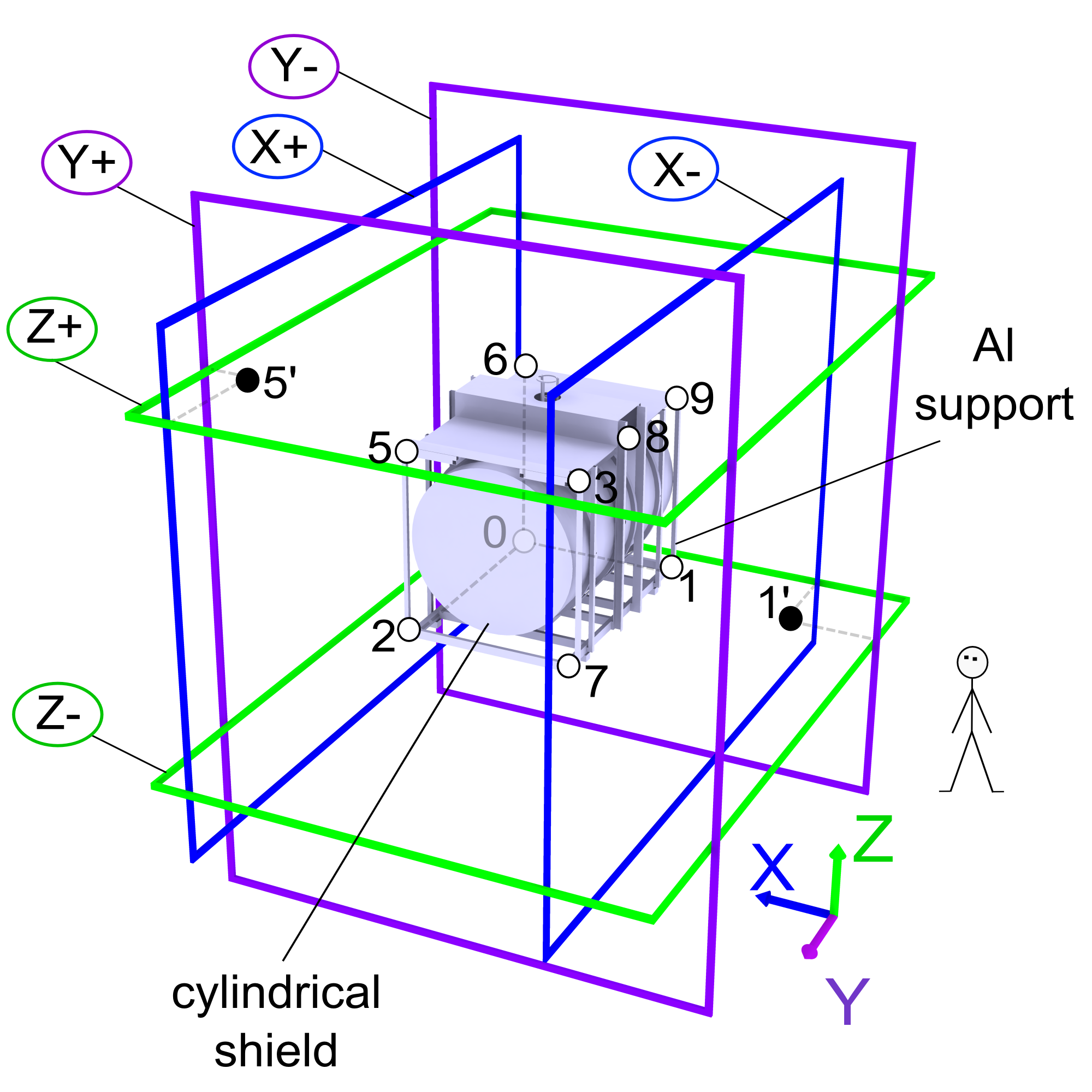}
  \caption{
Sketch of the SFC system consisting
of six coils surrounding the Mu-metal magnetic shield of the nEDM spectrometer.
The visible outermost layer of the cylindrical shield is mounted in its aluminum support structure.
The Helmholtz coil pairs are labeled (X$+$,X$-$),(Y$+$,Y$-$) and (Z$+$,Z$-$).
The coordinate system of the experiment is given at the lower right.
Its origin is at the center of the magnetic shield.
Three-axis fluxgates (open circles) are mounted on the Al support of the experiment 
and numbered according to the fluxgate nomenclature given in the text.
The positions $1'$ and $5'$ (full circles) depict previous locations of fluxgates FG\,1 and FG\,5 
referred to in Sec.\,\ref{sec:resultsWoMatrix}.
FG\,4 is omitted as it was removed from the system after a sensor failure.
	}
  \label{fig:SFCsetup}
\end{figure}  

 \begin{figure}
\centering
   \includegraphics[width=\OneColumnPictureWidth]{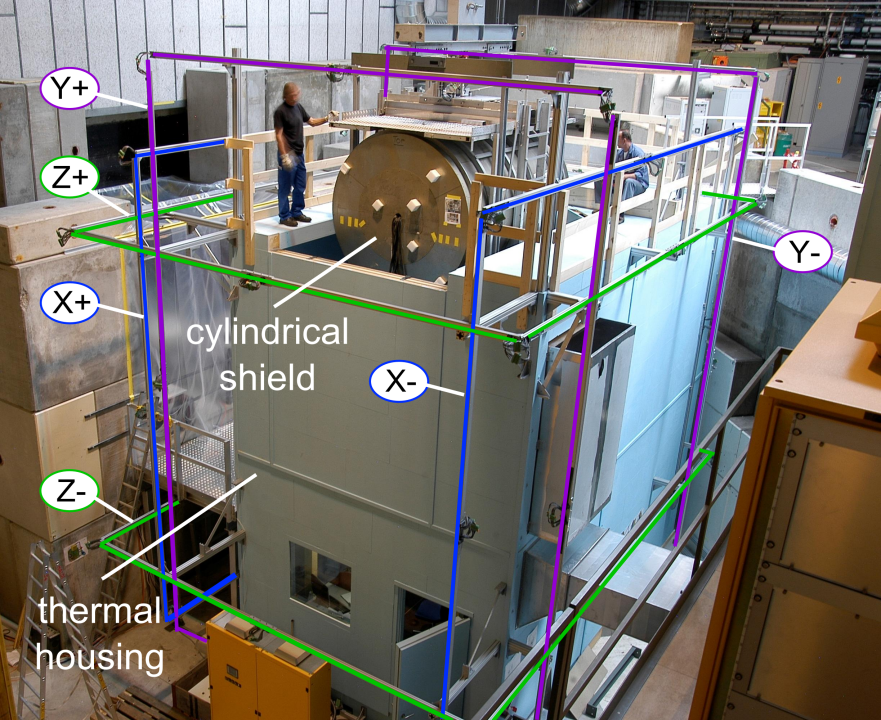}
  \caption{
	Photograph of the SFC system mounted around the temperature stabilized housing 
	-- better than 1$^\circ$~C around the magnetic shield --
	of the nEDM experiment.
  The picture was taken as the apparatus was being lowered into the housing in 2009.
  The outermost layer of the magnetic shield is visible.
  The biological shielding of the UCN source (concrete blocks) is visible in the back. }
  \label{fig:SFCfoto}
\end{figure}  

The six rectangular coils of the SFC,
labeled the (X$\pm$), (Y$\pm$), and (Z$\pm$)-coils,
consist of copper wires with 6\,mm$^2$ cross section, 
mounted on aluminum frames, 
which are designed out of electrically isolated and ungrounded bars.
They approximate a Helmholtz configuration 
as shown in Figs.\,\ref{fig:SFCsetup}
and \ref{fig:SFCfoto}.
Table \ref{tab:coils} summarizes important 
properties of the coils.

The origin of the experiment coordinate system is set at the center 
of the magnetic shield, which coincides with the center of the vacuum tank.
Each coil pair is centered at this origin as well as possible,
given physical constraints such as the presence of concrete blocks that are 
part of the biological shielding of the UCN source.
The maximum offset from the center along any of the axes is 
less than 0.2\,m.

The coils are driven by unipolar power supplies from FuG, 
type NTN350-35 and NTN700-35.
These can provide 350\,W and 700\,W  at a maximum voltage of 35\,V. 
We have These power supplies are specified to a relative accuracy of 0.2$\%$ by the manufacturer.
verified 
that the current can be controlled with 16-bit resolution. 
Usually the power supplies are operated at 70\,\% to 90\,\% of maximum current.
Software-controlled solid state relays were added to the system to allow 
change of coil polarity.
This was necessary because the superconducting test facilities 
SULTAN \cite{Bruzzone2002,Stepanov2013}
and EDIPO \cite{Portone2012},
at an approximate distance of 30\,m from our setup,
can cause a polarity change of the magnetic field in the horizontal plane at our 
experimental site during magnet ramping.

\begin{table}[htb]  
\begin{ruledtabular}
    \begin{tabular}{lcccccc}
Coil           &X$+$   &X$-$    &Y$+$    &Y$-$   &Z$+$  &Z$-$ \\ \hline
Long side (m)  &7.9  &7.9   &8.2   &8.4  &8.0 &8.0 \\ 
Short side (m) &6.1  &6.1   &6.0   &5.9  &5.8 &5.8 \\ 
Pair separation (m) & \multicolumn{2}{c}{4.2}&  \multicolumn{2}{c}{5.1} &  \multicolumn{2}{c}{4.2}\\ \hline
Windings              &18   &18    &9    &9    &12   &12  \\
Resistance ($\Omega$) &2.0  &2.0   &1.0  &1.0  &1.3  &1.3 \\
Inductance (mH)       &8    &8     &2    &2    &4    &4  \\
Static current (A)    &8.7  &8.7   &7.3  &7.3  &14.0 &14.0 \\ 
\end{tabular}
\end{ruledtabular}
\caption{\label{tab:coils}    
SFC coil dimensions, electromagnetic properties, and the static currents that 
are applied to the coils to compensate for the DC component of the surrounding magnetic field. 
The values for resistances and inductances 
were extracted from impedance measurements which were done at frequencies 
between 0.5\,Hz to 8\,kHz
with the complete nEDM setup installed.
}
\end{table}  

The surrounding magnetic field is monitored with ten 
three-axis fluxgate magnetometers from Bartington, 
type Mag-03 MCL70 or MCTP70,
mounted at the corners of the aluminum support structure of the Mu-metal 
shield, as shown in Fig.\,\ref{fig:SFCsetup}. 
The nomenclature of the sensors follows the pattern of fluxgate 
numbers \mbox{FG\,0-9} and their three orthogonal sensors in experiment 
coordinates: $\left\{ 0\mrm{x},0\mrm{y},0\mrm{z},
\ldots, 9\mrm{x},9\mrm{y},9\mrm{z}\right\}$.
Due to a sensor failure, fluxgate 4 (FG 4) was removed from the system.
However, the entire system and signal treatment 
was developed to accommodate up to 30 sensors.

The sensor signals are sampled with two 16-bit multiplexing 
analog-to-digital converters (ADC) 
at a rate of 15.45\,kHz per sensor. 
Before sampling, the signals are filtered by a passive low-pass filter with 43\,Hz bandwidth 
which was chosen to suppress aliasing at the least significant bit level.
The high sampling frequency, 
in combination with analog and digital filters,
provides a high amplitude resolution of the sensor signal.

The same filters also ensure that the feedback system 
operates at frequencies significantly lower
than the smallest magnetic resonance frequency in our system 
which is $\sim$8\,Hz from a $^{199}$Hg magnetometer \cite{PPnedmPSI,Baker2013}.
Under these conditions the data acquisition (DAQ) system has an 
internal noise floor of $\sim$10\,pT/$\sqrt{\mathrm{Hz}}$, 
which corresponds approximately to the intrinsic 
sensor noise.

The SFC control software has two operation modes: 
(i) static, where the currents in the coils are constant
and compensate the DC component of the ambient field;
the stability of the current output was measured to be at the 10$^{-5}$ level
with an ohmic resistance as load;
(ii) dynamic, where
a digital feedback loop 
monitors the magnetic field and 
controls the currents at an iteration rate of 6\,Hz,  
which is limited by the slow inherent settling time of the coil power supplies.

All relevant system properties are averaged if necessary 
and written to file at a rate of 1\,Hz.
%



\subsection{Magnetic field conditions for the nEDM measurement}
\label{sec:conditions}

Inside the Mu-metal shield of the nEDM experiment 
a cosine-theta coil wound around the cylindrical vacuum tank 
generates a vertical magnetic holding field of 1\,$\mu$T
necessary for the Ramsey method. 
Up to 33 trim coils can be used to homogenize the field to a level better than 10$^{-3}$. 
However, the holding field is superimposed by the remanent magnetic field  
of the Mu-metal shield.
In order to demagnetize the shield, a so-called idealization procedure 
\cite{Thiel2009,PTBdemag2008} is applied.
This procedure is analogous to a standard demagnetization but is done within a 
non-zero external and/or internal field and yields a reproducible remanent field of less than 1\,nT 
inside the vacuum tank measured 
over the region of interest,
in the case when the vertical holding field is turned off.
For this reason the shield is idealized at least once a day,
usually after a ramp of SULTAN or EDIPO.
Keeping the surrounding magnetic field stable
reduces the need for more frequent idealizations which would compete with
nEDM measurement time.
External perturbations can not only
influence the magnitude of the holding field, 
but also locally destabilize the magnetization state of the Mu-metal shield, 
which may then lead to
time-delayed changes of the Mu-metal magnetization.
This effect combined with the high shielding factor of the Mu-metal 
causes that often no direct correlation between external 
-- measured by fluxgates --
and internal -- measured by atomic magnetometers --
magnetic field noise is observable.
Thus, the control of the external magnetic field changes is done 
with the fluxgate sensors outside the magnetic shield, 
while the internal atomic magnetometers \cite{knowles}
are used to monitor the stability of the internal field.


\subsection{Magnetic field characteristics at the nEDM site}
\label{sec:fieldatsite}

The environmental surrounding magnetic field components in 
the experiment coordinate system are given by
\begin{equation} 
\boldsymbol{B}^\mathrm{surrounding} \begin{pmatrix} \mathrm{x}\\ \mathrm{y} \\\mathrm{z}\end{pmatrix} \approx
  \begin{pmatrix} +37\,\mu \mathrm{T} \\ +10\,\mu \mathrm{T} \\-41\,\mu \mathrm{T}\end{pmatrix} 
\label{eq:magfieldatorigin}
\end{equation} 
at the coordinate system origin, measured before the experiment and the Mu-metal 
shield were installed, and 
with SULTAN and EDIPO turned off.
The magnitude of this field is 
dominated by the Earth's magnetic field.
Additional static magnetic field contributions
originate from the typical environment 
at large research facilities,
e.g.~radiation shielding 
made of iron and concrete, steel columns of building walls, etc.
Gradients in the surrounding magnetic field and the distortion of the 
flux density due to the Mu-metal
cause absolute magnetic field values measured 
at single fluxgate positions to range up to to 85\,$\mu$T.

\begin{figure}
\centering
   \includegraphics[width=\OneColumnPictureWidth]{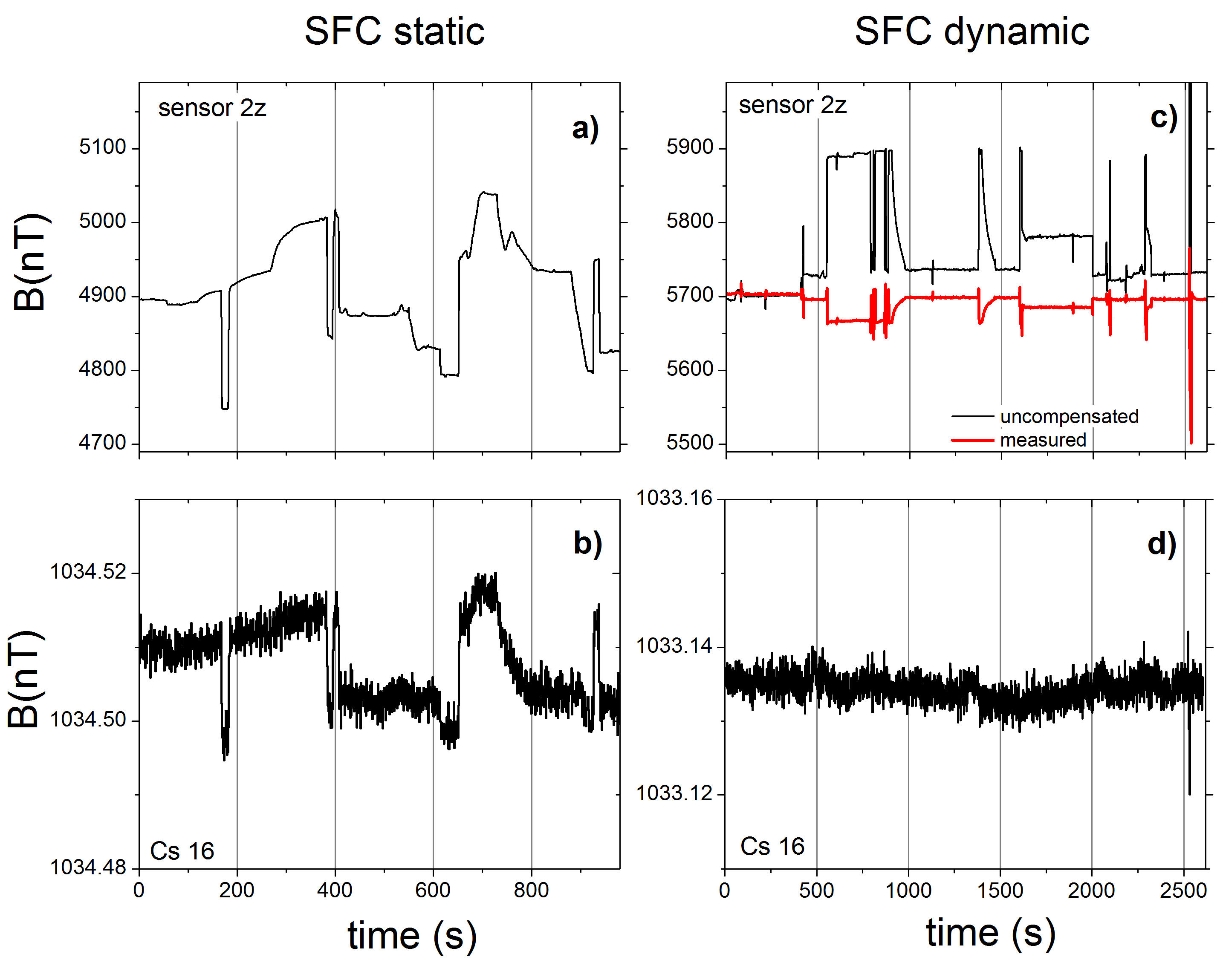}
  \caption{
Measurement of the magnetic field at the nEDM site 
during day-time using sensor 2z (a) and (c), 
and 
Cs magnetometer 16 inside the Mu-metal shield (b) and (d).\newline
a) The measured uncompensated field shows up to 300\,nT large variations;
b) the Cs reading with smaller variations at the same time, 
reflecting the shielding of the Mu-metal shield.
c)
SFC in dynamic mode: the measured compensated field (lower red curve) and 
the uncompensated field (upper black curve);
d)
the corresponding Cs reading showing that all field variations were 
smaller than the magnetometer noise. 
Only one large field spike of more than 250\,nT
-- 700\,nT in the uncompensated field --
at T=2530\,s is also observed inside the shield.
	}
  \label{fig:surrounding-field}
\end{figure}

Field perturbations during day-time occur rather frequently 
on a level of a few hundreds of nT.
Figs.~\ref{fig:surrounding-field} 
shows a $\sim$1000\,s long snapshot
of the field measured with a fluxgate sensor a) or a Cs magnetometer b).
Inside the Mu-metal shield
the variations are observed with highly sensitive cesium magnetometers \cite{knowles} 
operated inside in the 1\,$\mu$T magnetic holding field.
All ten installed Cs magnetometers show a similar behavior and observe a strongly reduced amplitude. 
This confirms the passive shielding factor of our Mu-metal shield
of order 10$^{4}$.

In order to demonstrate already here the power of our stabilization setup,
we plot for comparison a similar situation with the 
field compensation working in dynamic mode (see Sec.\ref{sec:Inplusone}).
Figs.~\ref{fig:surrounding-field}c) and d) 
reflect the situation for a standard day-time data taking period.
c) displays the measured field and the 
one calculated to be the original uncompensated field (see Eq.\ref{eqn:superpos}).
The suppression of the disturbances is obvious.
In d) the corresponding measurement with a Cs magnetometer demonstrates that 
only one large field spike is not compensated sufficiently and observed 
inside the shield.

\begin{figure}
\centering
   \includegraphics[width=\OneColumnPictureWidth]{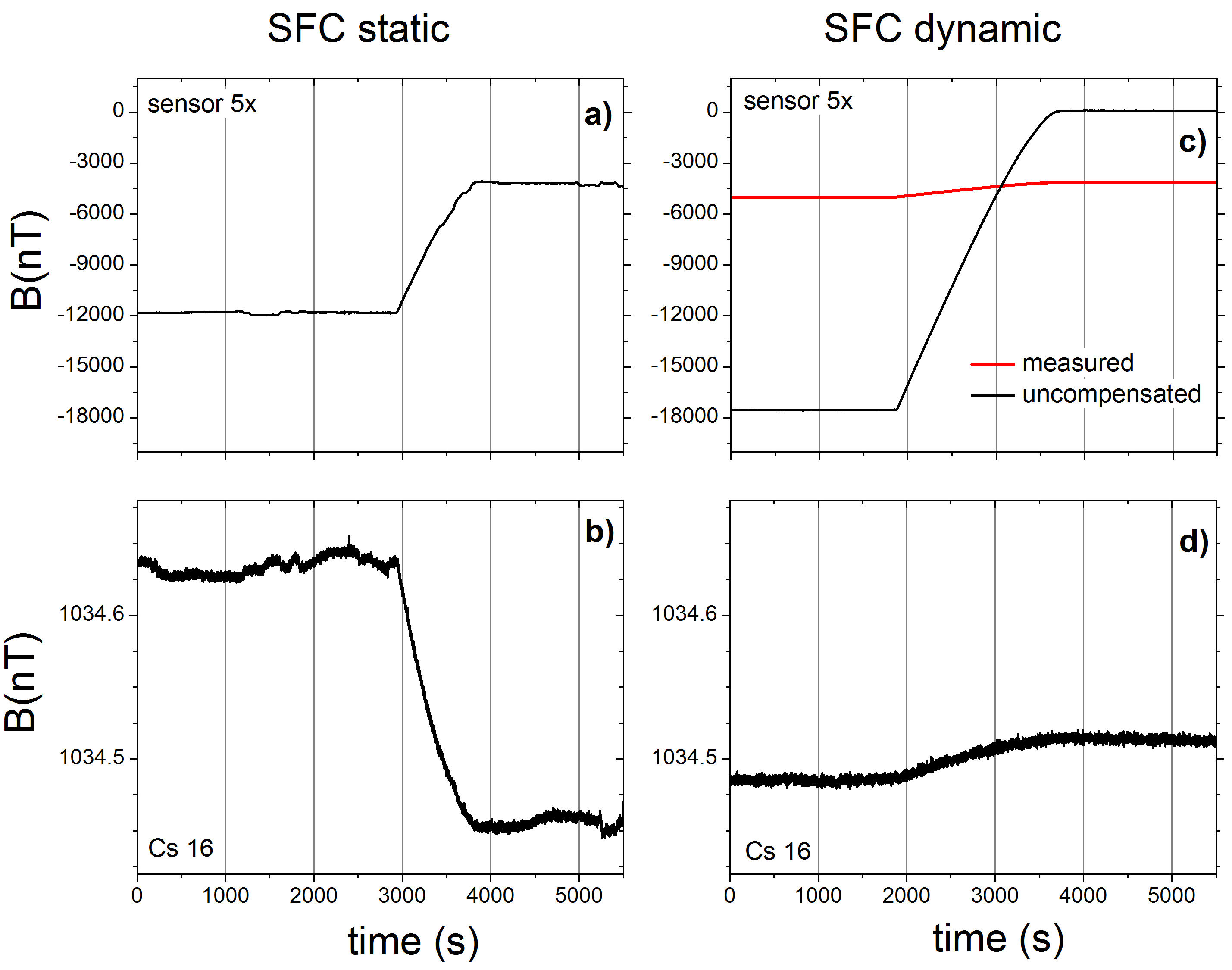}
  \caption{
Measurement of the magnetic field at the nEDM site 
during a field ramp of the SULTAN facility
using sensor 5x (a) and (c), 
and 
Cs magnetometer 16 inside the Mu-metal shield (b) and (d).
a) and b) display the observed field change without dynamic compensation,
c) and d) with the SFC in dynamic mode.
	}
  \label{fig:ramp}
\end{figure}

The largest disturbance at the nEDM site occurs during regular field ramps of 
the neighboring SULTAN facility, a situation where 
nEDM measurements are stopped.
Fig.~\ref{fig:ramp}a) shows the observed
$\sim$7.7\,$\mu$T field change caused by a 2.5\,T ramp
at the position of sensor 5x without dynamic field compensation.
At the same time the Cs magnetometer inside the Mu-metal shield 
observes a field change of about 180\,pT (Fig.~\ref{fig:ramp}b). 
The measured field change caused by a 5\,T ramp
with dynamic field compensation on 
is plotted in 
Figs.~\ref{fig:ramp}c) and d).
Outside the shield this disturbance is reduced by a shielding factor of about 20.
The ten available Cs magnetometers inside the shield
show corresponding shielding factors between 11 and 24.
Clearly we can state, 
that the suppression of disturbances by dynamic SFC and 
passive magnetic shield multiply.

The bandwidth of magnetic noise or perturbations which the compensation system is able to 
attenuate extends from 0.5\,Hz down to $\sim$$10^{-3}$\,Hz. 
This corresponds to time ranges of seconds to hours, which are 
the important time scales for the nEDM experiment.
Single nEDM measurements last from about 100\,s up to 300\,s,
and
measurement sequences for one parameter set take several hours.
Therefore, the magnetic field should be stable over such time scales.
Magnetic noise within this bandwidth is created at the site 
by neighboring experiments, passing vehicles, cranes, and other moving objects.

\begin{figure} 
\centering
  \includegraphics[width=\OneColumnPictureWidth] {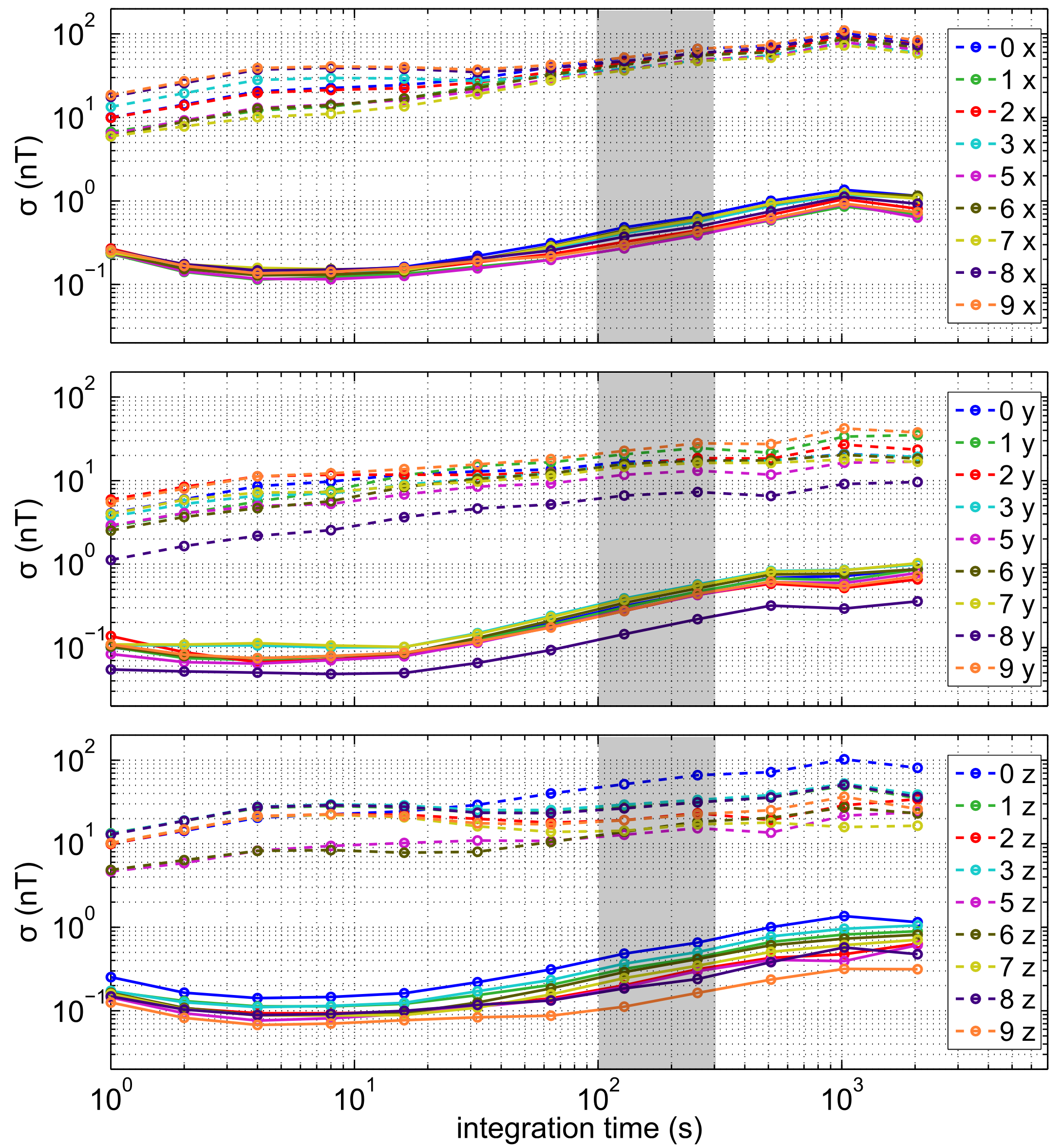}
  \caption{
  	Stability of the magnetic field during daytime (dashed lines) and 
	nighttime (solid lines) measurements, without dynamic stabilization shown
 	via the Allan 
	deviation $\sigma_{\mathrm{ADEV}}$ of all 
	SFC sensors. 
	Measurements are sorted by sensor number and orientation
	according to the coordinate system defined in Fig.\,\ref{fig:SFCsetup}: 
	$x$-sensors in the upper, $y$-sensors in the middle, 
	and $z$-sensors in the lower graph. 
	The gray area depicts the region of interest
	for the nEDM experiment.
}\label{fig:AllanNight}
\end{figure}

The stability of the magnetic field can be quantified via the Allan 
deviation 
$\sigma_{\mathrm{ADEV}}$  \cite{CharacOfFreqStab1971} which is a function of integration 
time $\tau$:
\begin{equation}
\sigma_{\mathrm{ADEV}}(\tau)=\sqrt{\frac{1}{2(N-1)}\sum^{N-1}_{l=1}\left( B^{(\tau)}_{l+1}-B^{(\tau)}_{l}\right)^2 } \hspace*{1mm},
\label{eqn:Allan0}
\end{equation}
where $N$=$T$/$\tau$ 
with $T$ being the total time of the measurement 
and $B^{(\tau)}_{l}$ the average magnetic field of the 
subsample $l$ over time $\tau$.
The Allan deviation of time $\tau$
is thus a measure of the average stability of the magnetic 
field at integration time $\tau$.
As shown in Fig.\,\ref{fig:AllanNight}, 
typical stability levels without dynamic stabilization
range from 10\,nT to 100\,nT 
during daytime -- without ramps -- and below 1\,nT
at nighttime and on weekends.

\subsection{Performance limitations}
\label{sec:multipoles}

The performance of the SFC is not only limited by its response bandwidth, 
but also by the number of adjustable coil currents,
resulting in six degrees of freedom.
A system of six coils generates fields dominated by constant terms 
and some first order gradients. 
In general, a first order gradient perturbation 
-- consisting of both constant and linear terms --  
contains eight independent parameters:
three homogeneous field components and five independent parameters of
the gradient tensor.
Thus, the compensation system is most effective at attenuating 
uniform magnetic field changes, e.g.~from magnetic field sources located far away.
Perturbations with their origin very close to, or even inside the SFC volume
can only be partially attenuated. 
Therefore, care must be taken to keep sources of magnetic noise away from the sensitive
volume of the experiment.


\section{Feedback algorithm for dynamic compensation}
\label{sec:Inplusone}

\subsection{Structure of the feedback algorithm}

\begin{table}
\begin{ruledtabular}
\begin{tabular}{lll}
Index &Quantity &Values\\
\hline
$n$  &Feedback iteration &$\left\{ \mathrm{1,2,3,\ldots} \right\}$\\
$j$  &Coil &$\left\{\mathrm{X+, X-,Y+,Y-, Z+, Z-}\right\}$\\
$k$  &Sensor &$\left\{ 0\mrm{x},0\mrm{y},0\mrm{z},\ldots, 9\mrm{x},9\mrm{y},9\mrm{z}\right\}$\\
$k'$ &Feedback sensor &Subset of $k$\\
\end{tabular}
\end{ruledtabular}
\caption{\label{tab:indices}	
	Summary of the indices for feedback iteration, coils,
	and sensors. 
}
\end{table}

In this work the currents $\boldsymbol{I}$ 
and magnetic field values $\boldsymbol{B}$ are 
summarized in arrays of size 6 and 27,
corresponding to the respective coils (index $j$) and field sensors (index $k$).
An array, as well as each array element, can have a superscript index $n$ 
referring to an iteration of the feedback loop, e.g.~$I_j^n$ 
is the current in coil $j$ in iteration $n$.
This indexing convention is summarized in Tab.~\ref{tab:indices}.

 \begin{figure}
\centering
   \includegraphics[width=\OneColumnPictureWidth]
   {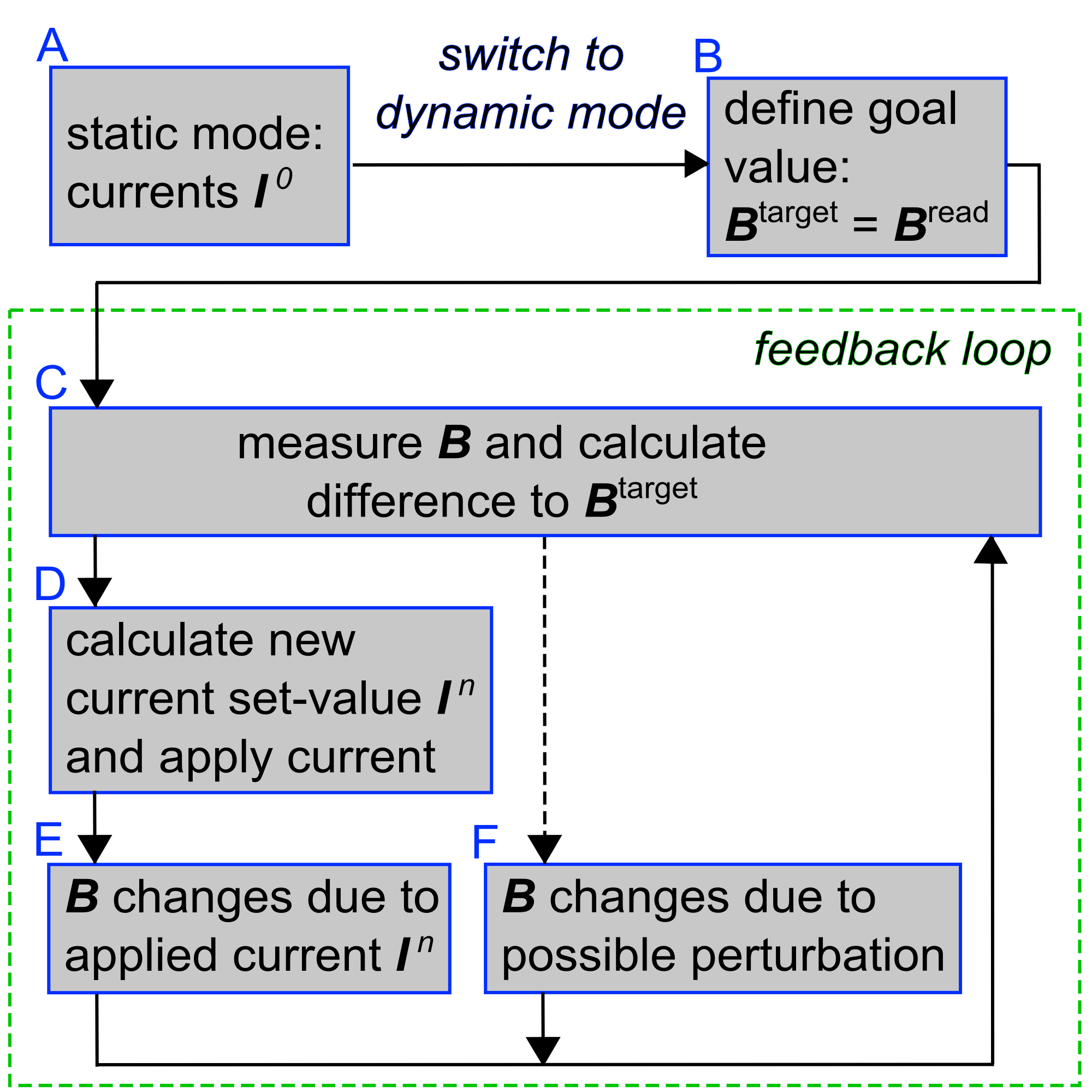}
  \caption{Flow chart of the structure of the SFC control algorithm. 
  The solid arrows indicate temporal 
	as well as causal sequences.
  In static mode, an array of six constant currents $\boldsymbol{I}^0$ 
	is applied to the SFC coils (step A).
  When switching into dynamic mode, the magnetic field 
	target-value $\boldsymbol{B}^\mathrm{target}$ is defined by the current 
	read-value $\boldsymbol{B}^\mathrm{read}$ and used in the feedback loop (step B).
  Then the feedback loop (enclosed by the green dashed rectangle) 
	is started and steps C to E are executed repeatedly.
  The dashed arrow from step C to step F indicates a temporal sequence, 
	but not a causal sequence.
  When the dynamic mode is stopped, the system goes back to static mode, 
	and the last current set-values $\boldsymbol{I}^N$ is used as new $\boldsymbol{I}^0$.}
  \label{fig:SFCfeedbackflow}
\end{figure}  

A flow chart of the main structure of the SFC control 
algorithm is given in Fig.\,\ref{fig:SFCfeedbackflow}.
In static mode, constant currents, $\boldsymbol{I}^0$, are applied to all coils.
During standard magnetic field conditions (SULTAN and EDIPO not in operation), 
the control system applies the currents given in Tab.~\ref{tab:coils}.
They partly compensate for the DC component of the environmental field and 
yield low absolute magnetic field values
(Sec.\,\ref{sec:DCshielding}).
The target-value of the magnetic field $\boldsymbol{B}^{\mathrm{target}}$ is 
not predefined within the feedback algorithm.
Instead, when switching from static to dynamic mode, 
the target-value $\boldsymbol{B}^\mathrm{target}$ is set to the actual 
read-value of the magnetic field $\boldsymbol{B}^\mathrm{read}$ at the instant of switching
to avoid sudden field changes.
Once magnetic field perturbations occur, the read-value $\boldsymbol{B}^\mathrm{read}$ 
will change and deviate from the target-value.
Within each iteration $n$ of the feedback loop, the difference 
between $\boldsymbol{B}^{\mathrm{read}}$ and $\boldsymbol{B}^{\mathrm{target}}$ should 
be reduced by determining an appropriate array of six new current 
set-values $\boldsymbol{I}^n$.

The distinctive feature of the
algorithm is a matrix of 
proportionality factors which correlate magnetic field changes at all sensor 
positions to current changes in the SFC coils.
This matrix (Sec.\,\ref{sec:matrixintro}) is
used to calculate the new current set-values in each iteration of the feedback 
loop (step D).
Before the matrix can be included as a constant 
into the feedback loop, 
it has to be inverted and regularized \textcolor{zema}{once}
(see Sec.\,\ref{sec:pinvreg}). 
Approaches containing a regularized or truncated pseudoinverse 
are also known from bio-magnetometry \cite{baumgarten2008,baumgarten2010}, 
where the sources of magnetic signals (e.g.~from magnetically targeted drugs) 
are derived from measured magnetic fields.
Another application is the localization of ferromagnetic objects 
buried in the ground \cite{haueisen2009,haueisen2010}.



\subsection{Calculating a new current set-value}

In an earlier version of the feedback algorithm each coil current was controlled 
individually to stabilize one sensor reading, 
i.e.~six sensors were used as feedback sensors.
This method had the drawback that the field was stabilized very 
well at the positions of the feedback sensors, 
but not
anywhere else within the SFC volume.
An example of such behavior is given in Fig.\,\ref{fig:Sact-farfeedback}.
Presently we employ 
a more advanced method which 
enables us to
use more than six feedback sensors 
and, thus, transfers the stabilizing effect of dynamic compensation from 
certain single points
to the requested control volume.



\subsubsection*{The matrix of proportionality factors}
\label{sec:matrixintro}

 \begin{figure*} 
\centering
   \includegraphics[width=1\textwidth]{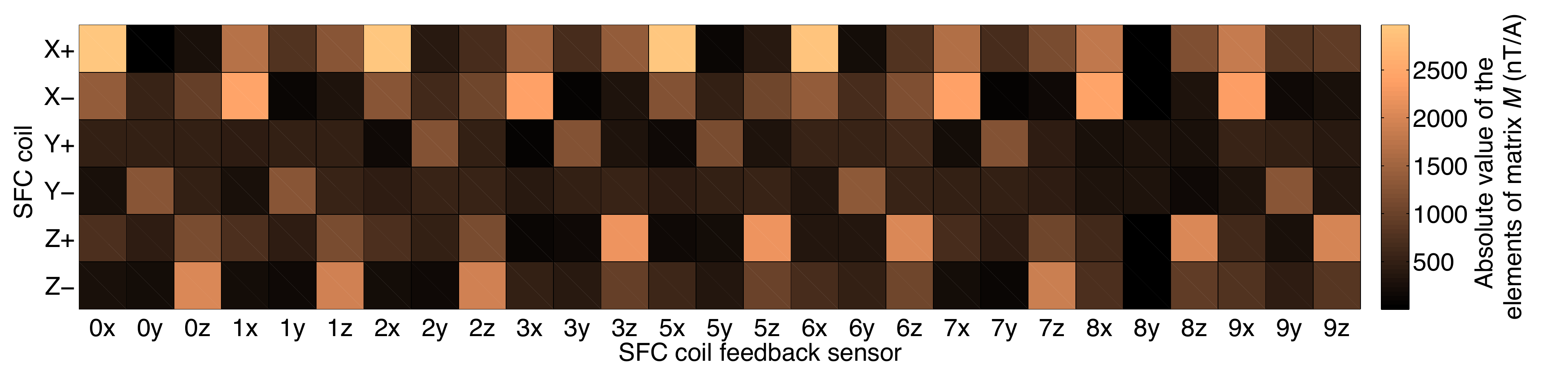}
  \caption{Color map of the absolute value of the elements of the matrix $M$.
  The SFC sensors are listed on the abscissa and the SFC coils on the ordinate.
  }
  \label{fig:SFCMatrix}
\end{figure*}  

We checked that 
each 
fluxgate sensor has a linear
response to current changes in each of the six SFC coils.
These proportionality factors (with units of nT/A) 
are summarized in a $27\times6$ 
matrix $M$ which is defined as:
\begin{equation}
B_k=\sum_jM_{kj}\cdot I_j   .
\label{eq:matrixdef}
\end{equation}

In the SFC setup the matrix elements, i.e.~the proportionality factors (hereafter used synonymously), 
vary by three orders of magnitude and reach values of up to a few 1000\,nT/A.
Their magnitudes depend on the orientation and specific position of the sensors, 
the distances to the SFC coils and the magnetic shield, and on irregularities of the Mu-metal, 
such as overlaps, feedthroughs, and welded joints.
A color map of the absolute values of the matrix elements $M_{kj}$
is shown in Fig.~\ref{fig:SFCMatrix}.
A pattern is recognizable and each sensor has the largest response to the particular coil
which corresponds best to its orientation and position.
For example, the largest matrix element of sensor 0x is $M(0\mathrm{x},\mathrm{X+})$ 
since it is aligned in the x-direction and mounted on the (X+)-side of the shield.



\subsubsection*{Including the proportionality matrix into a proportional-integral feedback algorithm}
\label{sec:PIfeedback}

In order to reduce the difference between 
$\boldsymbol{B}^{\mathrm{read}}$ and 
$\boldsymbol{B}^{\mathrm{target}}$ 
in a feedback loop, 
a compensating magnetic field has to be 
generated by modifying the coil currents.
This \textcolor{zema}{new} current is 
calculated by inverting Eq.\,\ref{eq:matrixdef}.
Since $M$ is not a square matrix, we use the 
Moore-Penrose pseudoinverse \cite{Moore1920,Penrose1955}
hereafter named pseudoinverse or $M^{-1}$.
Thus, when evaluating the change 
($\Delta \boldsymbol{I}$) to a new 
current set-value, all 27 available sensors 
$B_k$ can be taken into account:
\begin{equation}
\Delta I_j =  \sum_k M^{-1}_{jk} \cdot \left( B_k^{\mathrm{target}} - B_k^{\mathrm{read}} \right) =  \sum_k M^{-1}_{jk} \cdot \Delta B_k  ,
\label{eqn:bf_SFC_matrix}
\end{equation} 
where $\Delta B_k$ is the difference between the target-value and the read-value of sensor $k$.

To improve the stabilizing effect of dynamic SFC, 
the number of feedback sensors should theoretically 
be as high as possible.
However, using all available sensors as feedback 
sensors has the disadvantage that there are then 
no reference sensors left and no information about 
the magnetic field stability at non-stabilized 
points is available.
To avoid this, and to investigate the influence 
of the number of feedback sensors on the achieved 
stabilization, we tested the process with a subset 
of size $6<K'<27$.
Inserting Eq.\,\ref{eqn:bf_SFC_matrix} into a 
proportional-integral (PI) feedback algorithm yields 
the following formula for each current set-value at 
iteration $n$:
\begin{equation} 
I_{j}^n = I_{j}^0 \,\ + \underbrace{\vphantom{\sum_{n}}\alpha_j^{\mathcal{P}}\cdot \Delta I_{j}^n}_{\mrm{proportional\,\ term}}  + \,\ 
\underbrace{\alpha_j^{\mathcal{I}}\cdot \sum_{t=1}^{n} \Delta I_{j}^t}_{\mrm{integral\,\ term}}  ,
\label{eqn:bf_SFC_matrixPID}
\end{equation} 
where $I_{j}^0$ are the coil currents at the moment of 
switching into dynamic mode (Fig.\,\ref{fig:SFCfeedbackflow}, step A).
The compensation can be tuned individually with the proportional and 
integral gains $\alpha_j^{\mathcal{P}}$ and $\alpha_j^{\mathcal{I}}$ 
for each coil $j$. 
$\Delta I_{j}^n$ is given by
\begin{equation} 
\Delta I_{j}^n = \sum_{k'} \hat{M}^{-1}_{jk'} \cdot \Delta B_{k'}^n   ,
\label{eqn:bf_SFC_DeltaI_n}
\end{equation} 
where $\hat{M}^{-1}$ is the pseudoinverse of a submatrix of $M$ which 
contains only the proportionality factors $M_{k'j}$ of the chosen $K'$ sensors used for the feedback.
The dimension of $\hat{M}^{-1}$ is thus $6\times K'$.


\subsection{Matrix inversion and regularization}
\label{sec:pinvreg}

The pseudoinverse of a matrix $M$ is calculated via the singular value 
decomposition \cite{PseudoInv1965}:
\begin{equation} 
M=U\cdot V \cdot W^{\mathrm{T}}\,\ \Rightarrow  \,\ M^{-1}=W\cdot V^{-1} \cdot U^{\mathrm{T}}  ,
\label{eqn:svd} \end{equation} 
where $U$ and $W$ are unitary matrices and
$V$ is a real diagonal matrix of the same dimensions as $M$, 
which contains the singular values $v_j$ of $M$.

Tests showed that applying the method described so far 
yields unsatisfactory results for our feedback system
as the stability of the magnetic field decreased in
the dynamic mode.
This can be explained by the large differences in magnitude
between the individual
matrix elements $M_{kj}$, i.e.~the matrix is ill-conditioned.
As a consequence, sensors with smaller matrix elements $M_{kj}$
have larger weights after matrix inversion.
Noise on one of these sensors will then be overcompensated 
and lead to instabilities.
Such effects are accompanied by large amplitudes in the current 
change $\Delta I_{j}^n$ during dynamic stabilization.
In order to avoid such noise amplification, a regularization is applied to the inversion, 
in our case a Tikhonov regularization \cite{Tikhonov1977}.
This method replaces 
the inverted singular values $v_j^{-1}$ of the matrix in the following way:

\begin{equation} 
V^{-1}_{jj}=\frac{1}{v_j} \,\ \rightarrow \,\ \frac{v_j}{v^2_j+\beta^2}  ,
\label{eqn:tikhonov} 
\end{equation}
where $\beta = 10^r\,\mathrm{nT/A}$ and $r$ is the regularization 
parameter with a range of $-\infty <r<\infty$.

The limit $r\rightarrow-\infty$ corresponds to the non-regularized 
pseudoinverse of the matrix $M$.
Setting $r$ to $+\infty$ 
will result in $V_{jj}^{-1}\rightarrow 0$, 
and, from 
Eq.\,\ref{eqn:svd},
it will also result that $M^{-1}\rightarrow 0$.
The regularization 
has the greatest effect 
when $\beta$ is of the order of $v_j$, 
which corresponds to approximately \mbox{$2\leq r \leq 4$}
in our case.

%

\subsubsection*{Determination of the regularization parameter}

 \begin{figure*} 
\centering
   \includegraphics[width=1\textwidth]   {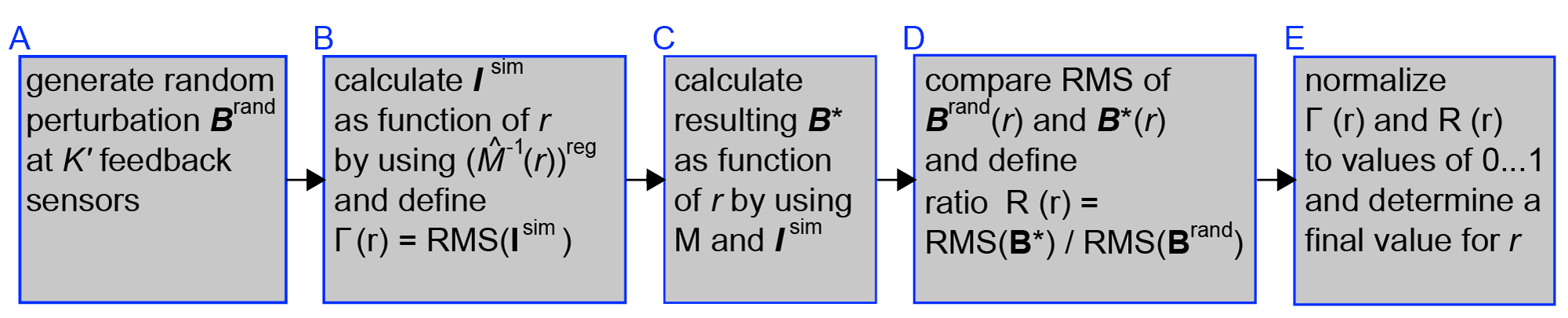}
  \caption{Flow-chart of the process used to optimize the regularization parameter.}
  \label{fig:toymodel}
\end{figure*}  

We simulated 
a simplified compensation model 
which includes the measured proportionality factors 
$M_{kj}$ in order to choose an appropriate value of $r$.
A flow chart of the concept is shown
in Fig.~\ref{fig:toymodel}.

First (step A in Fig.~\ref{fig:toymodel}), 
for the chosen number of feedback sensors
$K'$ magnetic field 
values are 
picked at random, from a normal distribution that is 
based on the noise spectrum observed at the apparatus 
in the bandwidth of interest, to form 
$\boldsymbol{B}^\mathrm{rand}=\{B_1^\mathrm{rand},B_2^\mathrm{rand},...,B_{K'}^\mathrm{rand}\}$.

Next (step B), an array of simulated current set-values 
$\boldsymbol{I}^\mathrm{sim}$ 
is calculated 
using Eq.\,\ref{eqn:bf_SFC_DeltaI_n}:
\begin{equation} 
I_{j}^\mathrm{sim}\left( r\right)=\sum_{k'} 
\left(\hat{M}^{-1}_{jk'}(r)\right)^\mathrm{reg}\cdot \left(-B_{k'}^\mathrm{rand}\right)  .
\label{eqn:currentcommand_sim} 
\end{equation} 
$\boldsymbol{I}^\mathrm{sim}(r)$ varies not only as a function of $r$, 
but also as a function of $\boldsymbol{B}^\mathrm{rand}$.
Thus many different $\boldsymbol{B}^\mathrm{rand}$ have to be compared 
in order to determine 
how much the response of our feedback can vary.
Typical field change distributions observed in the real system 
are shown in Appendix E5 of 
Ref.~\cite{DissBea}.

 \begin{figure} 
\centering
 \includegraphics[height=\OneColumnPictureWidth, angle=90]{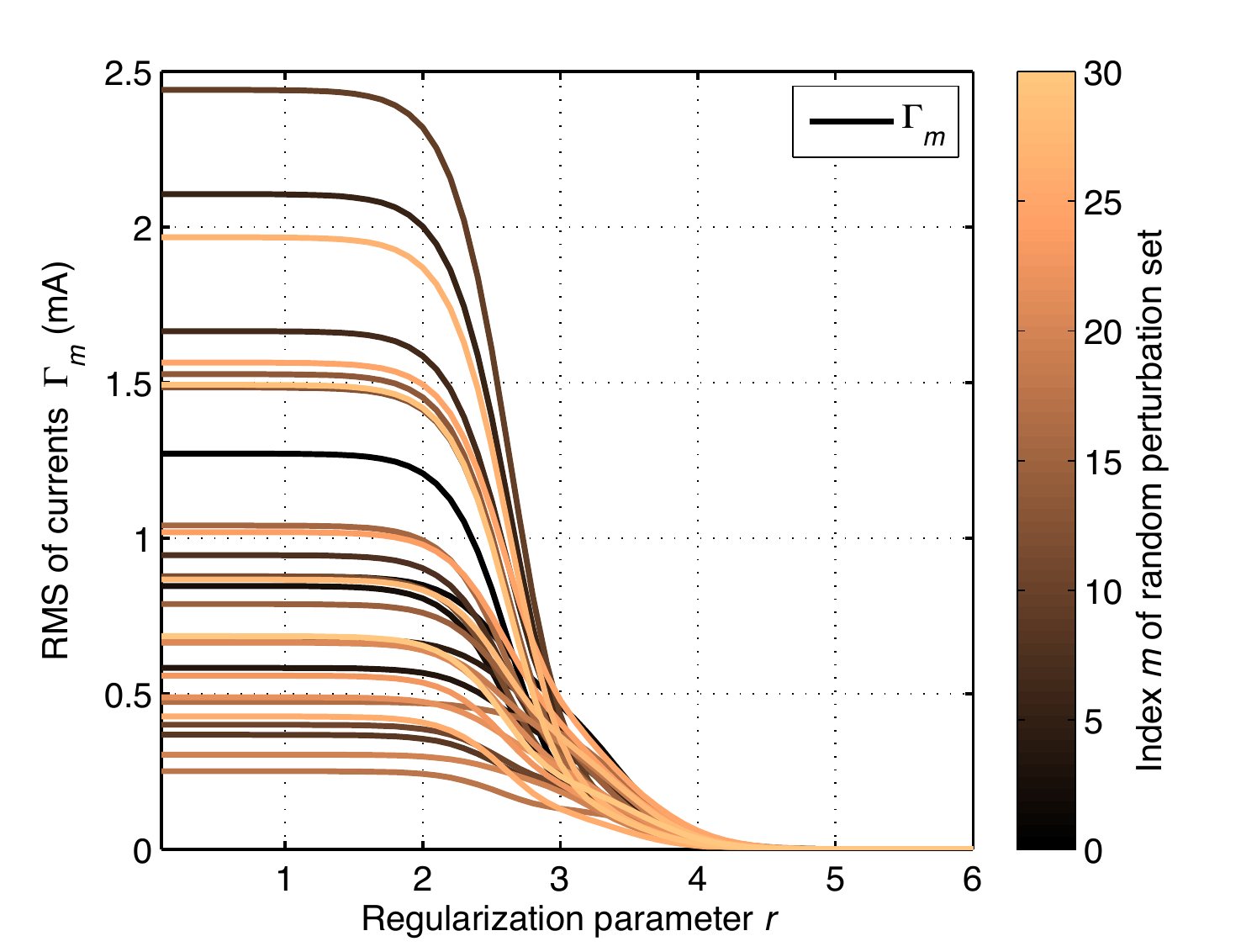}
  \caption{RMS of simulated currents $\Gamma_m$ versus regularization 
	parameter $r$ for 30 different sets $m$=1...30
	of random perturbations $\boldsymbol{B}^\mathrm{rand}$. 
	For details see text.
	} \label{fig:currentRMS} 
\end{figure} 

As a figure of merit, the root mean square (RMS) of the currents, 
$\Gamma$, is defined as a function of $r$:
\begin{equation} 
 \Gamma(r)=\sqrt{\frac{1}{6}\, \sum_{j=1}^6 (I_{j}^\mathrm{sim}(r))^2}  .
\label{eqn:currentcommand_rms} \end{equation} 
In the following example we used 30 different sets of 
$\boldsymbol{B}^\mathrm{rand}$
indexed with $m=1...30$.
The 30 resulting $\Gamma_m(r)$ for the 
different $\boldsymbol{B}^\mathrm{rand}$ are 
plotted in Fig.\,\ref{fig:currentRMS} with
each $m$ indicated in a different color.
With increasing regularization parameter, 
the magnitude of the currents in the SFC coils decreases. 
Lower compensation currents (not the DC part though) indicate smaller overcompensation,
but if currents are too small the stabilizing effect will vanish;
hence a compromise has to be found.
The resulting magnetic field $\boldsymbol{B}^*$ is 
a superposition of the perturbation 
$\boldsymbol{B}^\mathrm{rand}$ and the field 
caused by $\boldsymbol{I}^\mathrm{sim}$.
This field can be calculated with the original 
submatrix $\hat{M}$ of the feedback sensors $K'$ 
(step C).
Thus $\boldsymbol{B}^*$ as function of $r$ is given by: 
 \begin{equation} 
  B_{k'}^*(r)=B_{k'}^\mathrm{rand}+\sum_j\hat{M}_{k'j}\cdot I_{j}^\mathrm{sim}(r)  .
\label{eqn:resultingB} \end{equation} 

In order to evaluate the effect of the simulated currents on the field 
perturbation, we compare the RMS of $B_{k'}^\mathrm{rand}$ 
to the RMS of the resulting compensated field $B_{k'}^*$ 
(step D),
\begin{equation} 
b=\sqrt{\frac{1}{K'}\, \sum_{k'=1}^{K'} \left(B_{k'}^{\mathrm{rand}}\right)^2}\,\,\,
\mathrm{and}\,\ b^*=\sqrt{\frac{1}{K'}\, \sum_{k'=1}^{K'} \left(B_{k'}^*\right)^2}  ,
\label{eqn:perturbationRMS} \end{equation} 
via the
ratio $R = b^*/b$. 

\begin{figure}
\centering
 \includegraphics[width=\OneColumnPictureWidth]{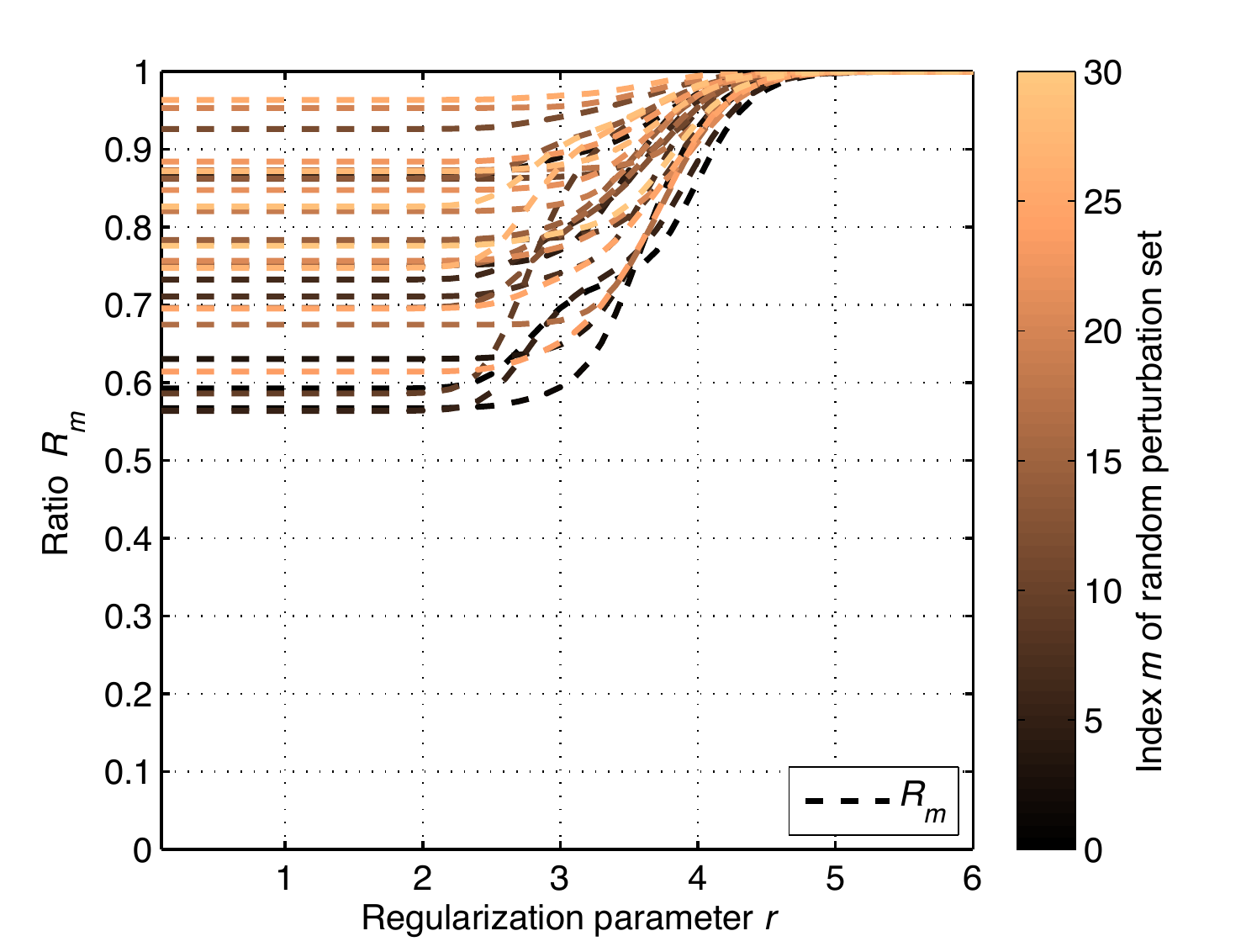}
  \caption{Ratio $R_m$ plotted versus regularization 
	parameter $r$ for 30 different sets of random perturbations $\boldsymbol{B}^\mathrm{rand}$. 
	Plot description as in Fig.\,\ref{fig:currentRMS}; the result of the same 30 
	perturbation sets is shown.} 
\label{fig:achievablecomp} 
\end{figure}

If the perturbation was compensated completely, 
the resulting field, and the ratio $R$ would be zero.
The dependence of $R$ 
for the same 30 values of $\boldsymbol{B}^\mathrm{rand}$
on the regularization parameter $r$ is shown in Fig.\,\ref{fig:achievablecomp}.
One can see that if the regularization parameter $r$ is too large, 
the compensation effect collapses completely 
as a consequence of the vanishing currents.
This figure also shows that in this simulated case, 
perturbations can only  be compensated for by 
a maximum of 45\%, 
a behavior also observed in the real system.

\begin{figure} 
\centering
  \includegraphics[width=\OneColumnPictureWidth,angle=0]{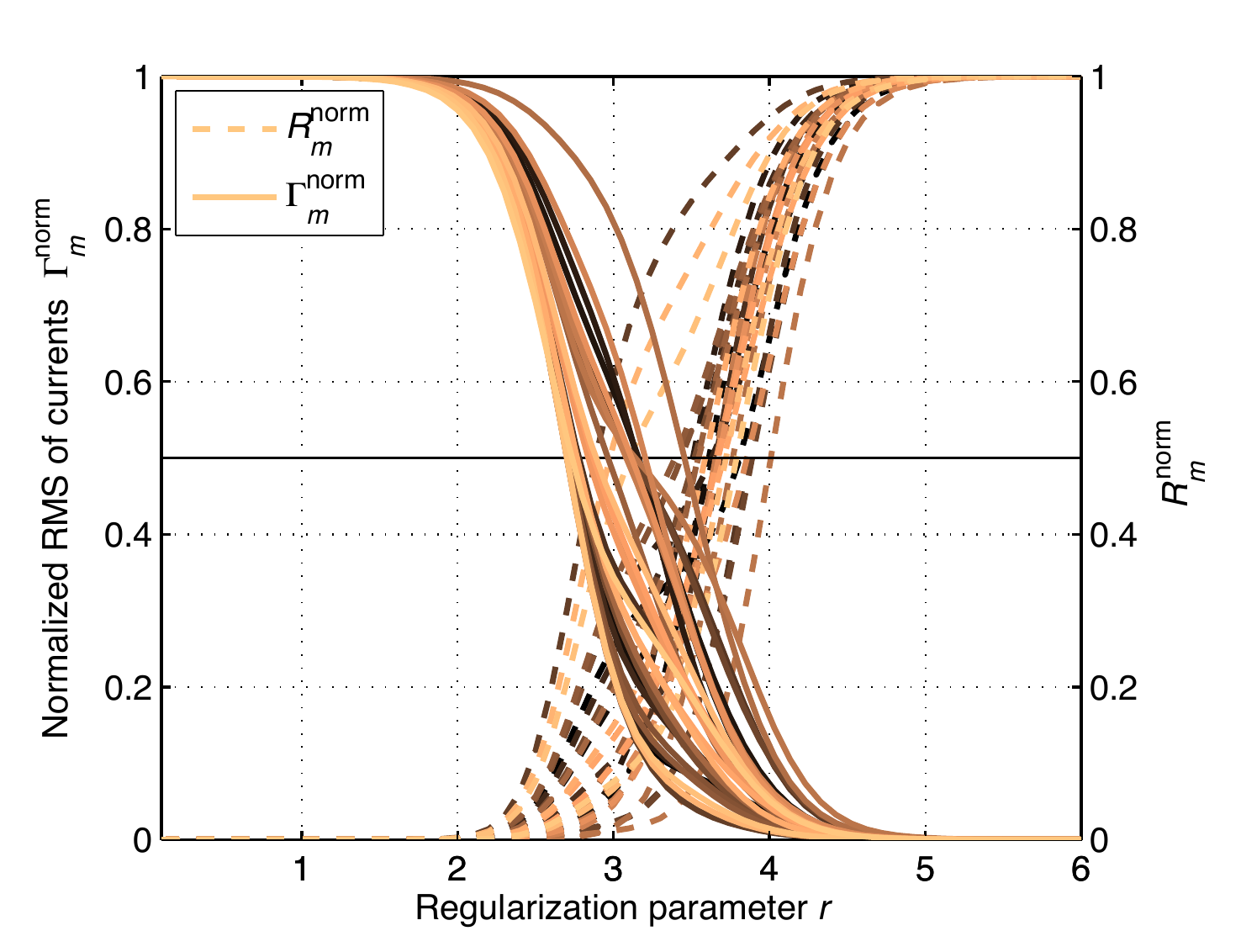} 
  \caption{Combined plot of $\Gamma_{m}^\mathrm{norm}$ and $R_{m}^\mathrm{norm}$ 
	versus regularization parameter $r$ for the same 30 perturbation sets shown in 
	Figs.\,\ref{fig:currentRMS}~and~\ref{fig:achievablecomp}. The same color code applies. 
	The average over the values at $\Gamma_{m}^\mathrm{norm}=R_{m}^\mathrm{norm}=0.5$ 
	determines the final choice of $r$.
	}\label{fig:regparchoice}
\end{figure}

In order to find the optimal value for $r$,
it is convenient to normalize $\Gamma$ and $R$ 
such that their minimum and maximum values 
lie between 0 and 1.
The resulting quantities  
$\Gamma_{m}^\mathrm{norm}(r)$ and 
$R_{m}^\mathrm{norm}(r)$ 
are shown in Fig.~\ref{fig:regparchoice}. 
The regularization parameter $r$ is finally determined in the 
following way (step E):
1000 different random values of $\boldsymbol{B}^\mathrm{rand}$ are generated.
For each random perturbation the values of $r$ 
corresponding to $\Gamma_{m}^\mathrm{norm}=0.5$ 
and to $R_{m}^\mathrm{norm}=0.5$ 
(as indicated in Fig.\,\ref{fig:regparchoice} by 
the horizontal solid grid-line) are stored.
These 2000 values of $r$ are then averaged.
In the model example shown here, the 
\textcolor{zema2}{final result 
of the described procedure is $r_\mathrm{final}=3.3$.}
The regularized pseudoinverse resulting from 
$r_\mathrm{final}$ is then fixed in the feedback algorithm.

A further fine tuning of the feedback behavior is possible via 
$\alpha^\mathcal{P}$ and $\alpha^\mathcal{I}$. 
Using different amplitudes and widths for the 
normal distribution from which we extract
$\boldsymbol{B}^\mathrm{rand}$,
we have found that the qualitative behavior of 
$\Gamma_{m}^\mathrm{norm}$ and $R_{m}^\mathrm{norm}$ 
does not depend on the magnitude of the perturbation.
The specific dependence on the regularization parameter $r$ is rather 
a property of the matrix itself, i.e.~of the particular subset of chosen feedback sensors.
Each time a different set of feedback sensors is used, $r$ is
re-determined in the way described above and the resulting 
regularized pseudoinverse is inserted into the feedback algorithm.

In general, we can compare tuning 
$\alpha^\mathcal{P, I}$
to changing $r$: in Fig.\,\ref{fig:regparchoice}
one can see that close to the chosen 
value of the regularization parameter, 
the dependency of the current on $r$ 
can be approximated to be linear.
In this way, the final fine tuning of 
the system can be achieved without 
recalculating a new regularized matrix.
$\alpha^\mathcal{P, I}$ are finally chosen to achieve a fast system response
and avoid overshooting.
Further detailed information about gain tuning can be found in 
Ref.~\cite{DissBea}.


\section{A method to quantify the SFC performance}
\label{sec:shieldingfactor}

\subsection{General comments on shielding factors}

There exist many different approaches to quantifying
the performance of an active stabilization system for magnetic fields, 
depending on each specific setup.
Often shielding factors include the passive shielding provided
by a high permeability magnetic shield, such as our Mu-metal shield,
which is usually one order of magnitude per shielding layer at 
frequencies around 1\,Hz and increasing for higher frequencies, as e.g.~in \cite{TerBrake1993}.
It is also common to investigate the shielding performance with artificial noise sources 
such as dedicated coils \cite{Kobayashi20124554}.
Such tests have limited significance since the coils are often of the same geometry as the 
compensation coils and thus generate a similar field. 
Furthermore, the excitation coils are typically
mounted with their axes parallel to one of the axes of the compensation setup.
As a consequence, the shielding factors are only determined for one direction \cite{PSIMagstab2005}, 
neglecting the fact that one-dimensional corrections
can increase the noise level in orthogonal directions.
This is referred to as cross-axial interference in \cite{Kobayashi20124554}.
Realistic shielding factors for arbitrary environmental magnetic noise are usually much smaller.
They are also more difficult to estimate, since random noise, 
as typically present in an accelerator facility such as ours, cannot  be easily reproduced.
The method presented here allows determining the shielding factors of all sensors in 
a single measurement and
does not rely on comparing measurements with compensation ``on'' vs. ``off''.


\subsection{Definition of the shielding factor}
\label{sec:ourshieldingfactor}
To investigate the impact of the SFC on the stability of the magnetic field, 
the measured field is interpreted as a superposition of the uncompensated 
field and the additional magnetic field created by the SFC coils at all sensor positions:
\begin{equation}
B_{k}^\mrm{meas}=B_{k}^\mrm{uncomp}+B_{k}^\mrm{coils}.
\label{eqn:superpos}
\end{equation}
The field generated by the coils can be calculated via
\begin{equation}
B_{k}^\mrm{coils}=\sum_jM_{kj}\cdot I_j    ,
\label{eqn:Bcoils}
\end{equation} 
and the magnetic field without the compensation effect $B^{\mrm{uncomp}}$ can be extracted.
The applied current values $I_j$ as measured by the coil power supplies are used. 
The Allan deviation (Eq.~\ref{eqn:Allan0}) is used as a measure of the magnetic field stability.
Comparing $\sigma_{\mathrm{ADEV}}$ of $B^{\mrm{uncomp}}$ to that of the measured 
magnetic field $B^{\mrm{meas}}$ reveals if the noise level is 
decreased by applying the SFC in dynamic mode.
The ratio of both $\sigma_{\mathrm{ADEV}}$ shows the factor by which the stability was improved at a 
given integration time $\tau$ for each specific measurement, independent of any reproducibility 
of the surrounding magnetic field properties.
The active shielding factor $S_k$ for each sensor is thus defined as
\begin{equation}
S_k(\tau)=\frac{\sigma_{\mrm{ADEV}}(B_{k}^\mrm{uncomp})}{\sigma_{\mrm{ADEV}}(B_{k}^\mrm{meas})},
\label{eqn:Sact}
\end{equation} 
regardless whether $k$ is a feedback sensor or not.
$S_k$ therefore reflects the improvement under real environmental conditions.

The largest possible integration time $\tau$ for a time series of length $T$ is 
$T/2$.
For computational reasons we  calculate the shielding factor only for $\tau=1,2,4,...,2^n\,$s.
Thus, each
time series under consideration 
is truncated to $T'=2^{n,\mathrm{max}}\leq T$. 
In this way the same portion of the time series is regarded for each $\tau$.
$S_k(\tau_\mathrm{max})$ is omitted in the plots
shown in Sec.\,\ref{sec:results}. 
The length of the measured time series was at least four times the largest given 
integration time.

A measurement with SFC in static mode, 
where the DC component of the environmental field is compensated, 
results in 
$S$ = 1.00000$\pm$0.00001  
for all integration times $\tau$ and therefore 
confirms the validity of the shielding factor definition. 
This demonstrates that including the coil currents in the shielding factor calculation 
(Eq.~\ref{eqn:Bcoils}) does not significantly affect the shielding factor values.

The statistical errors on the level of single magnetic field measurements 
have a negligible contribution to the shielding factor. 
The observed spread of shielding factors at various positions 
(plotted for several conditions in Sec.~\ref{sec:results}) 
represents in our experience a measure for the uncertainty 
of the shielding factor.


\section{Performance of the SFC system}
\label{sec:results} 
 
\subsection{Reduction of the DC component of the magnetic field around the Mu-metal shield}
\label{sec:DCshielding}

The amplitudes of magnetic perturbations are usually 
much smaller than the absolute value of the surrounding 
magnetic field given in Eq.\,\ref{eq:magfieldatorigin}. 
Thus, for the static operation mode,
a set of standardized currents was derived (Tab.~\ref{tab:coils})
which decreases the absolute value of the surrounding 
magnetic field to at least below
10\,$\mu$T at all sensor positions
but exceptions up to 15\,$\mu$T may occur at two sensor positions. 
These standard currents are also set during the demagnetization 
procedure of the shield.
The average DC reduction factor $\frac{1}{k}\sum_k|B_k^\mathrm{uncomp}|/|B_k^\mathrm{meas}|$ is 
about 20 or larger.


\subsection{Performance of the SFC with six independent fluxgate sensors}
\label{sec:resultsWoMatrix} 

In a simple feedback mode, six sensors are used to control the six SFC coil 
currents as listed in Tab.\,\ref{tab:feedbackSensorsPlain} via six independent control loops.

\begin{table}
\begin{ruledtabular}
\begin{tabular}{lcccccc}
Coil &X$+$ &X$-$ &Y$+$ &Y$-$ &Z$+$ &Z$-$\\
\hline
Feedback sensor         &5x   &1x   &5y   &1y   &5z   &1z\\
$\alpha^\mathcal{P}_j$  &0.95 &0.92 &0.98 &0.94 &0.45 &0.50 \\ 
 $\alpha^\mathcal{I}_j$ &0.44 &0.45 &0.42 &0.39 &0.35 &0.32\\
\end{tabular}
\end{ruledtabular}
\caption{\label{tab:feedbackSensorsPlain}    
Conditions of a simple feedback mode 
with six independent sensors 
and their proportional and integral gains, 
$\alpha^\mathcal{P}_j$ and 
$\alpha^\mathcal{I}_j$.
}
\end{table}

\begin{figure}  
\centering
\includegraphics[width=\OneColumnPictureWidth]{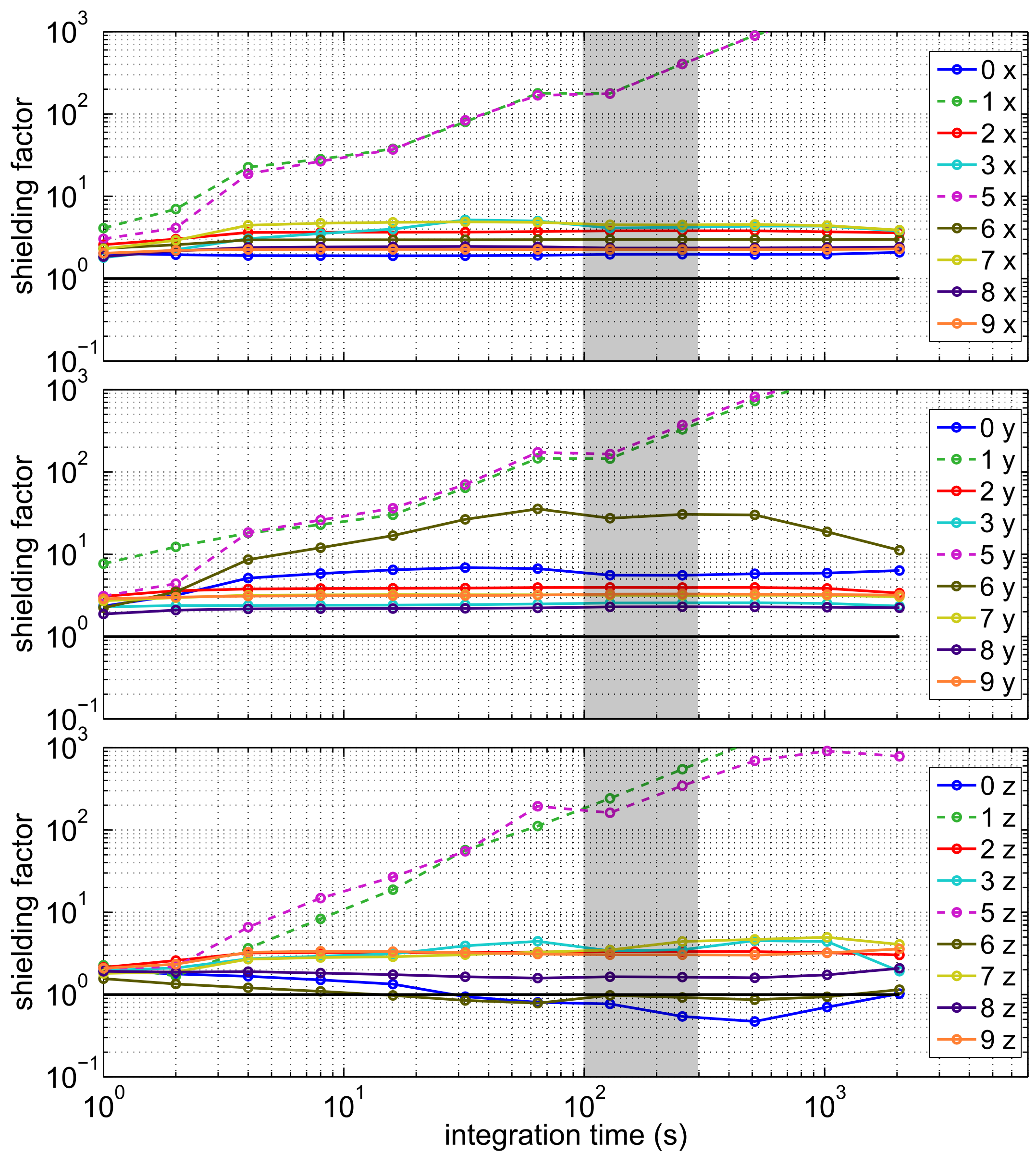}
\caption{Shielding factors $S_k$ for a measurement in simple six-sensor feedback mode.
The plot shows $S_k$ for all SFC sensors vs.~integration time, sorted by their orientation:
$x$-sensors in the upper, $y$-sensors in the middle, and $z$-sensors in the lower graph, respectively. 
Feedback sensors are plotted with dashed lines, monitoring sensors are plotted with solid lines. 
The solid black line is an emphasized gridline at $S_k$=1; shielding factors lower than one indicate 
noise increase by dynamic SFC implementation.
The gray area depicts the region of interest
for the nEDM experiment.
}
\label{fig:Sact-farfeedback}
\end{figure}

The fluxgate sensors FG\,1(x,y,z) and FG\,5(x,y,z) 
were located at positions $1'$ 
and $5'$ (full circles in Fig.\,\ref{fig:SFCsetup}).
Each position is at the crossing point of the three coil planes.
There, the smallest value for the
response of fluxgate sensors orthogonal to the axis of a 
given coil was found.
The signal was about 10\,\% of the sensor parallel to the coil axis.
The stabilization worked well only at the positions 
of the feedback sensors.
The stabilizing effect showed a huge discrepancy 
between the feedback sensors and all other sensors, as shown 
in Fig.\,\ref{fig:Sact-farfeedback}.
The shielding factor 
for the feedback sensors (dashed lines) reached values 
up to 10$^3$ and above for $\tau\geq1000\,$s, while other sensors 
showed low values from 2 to 4 or even indicated a decrease 
in stability in dynamic mode, e.g.~$S_{0z}<1$, 
(Fig.\,\ref{fig:Sact-farfeedback}, bottom).
Such a decrease of stability was even more prominent 
in magnetically quiet times
when very small perturbations at the positions 
of FG\,1 or FG\,5 were overcompensated and projected onto the entire control volume.

\subsection{Comparison of the SFC performance with twelve feedback sensors 
with a non-regularized and a regularized matrix.}

\label{sec:results12}
\begin{figure} 
\centering
\includegraphics[width=\OneColumnPictureWidth] {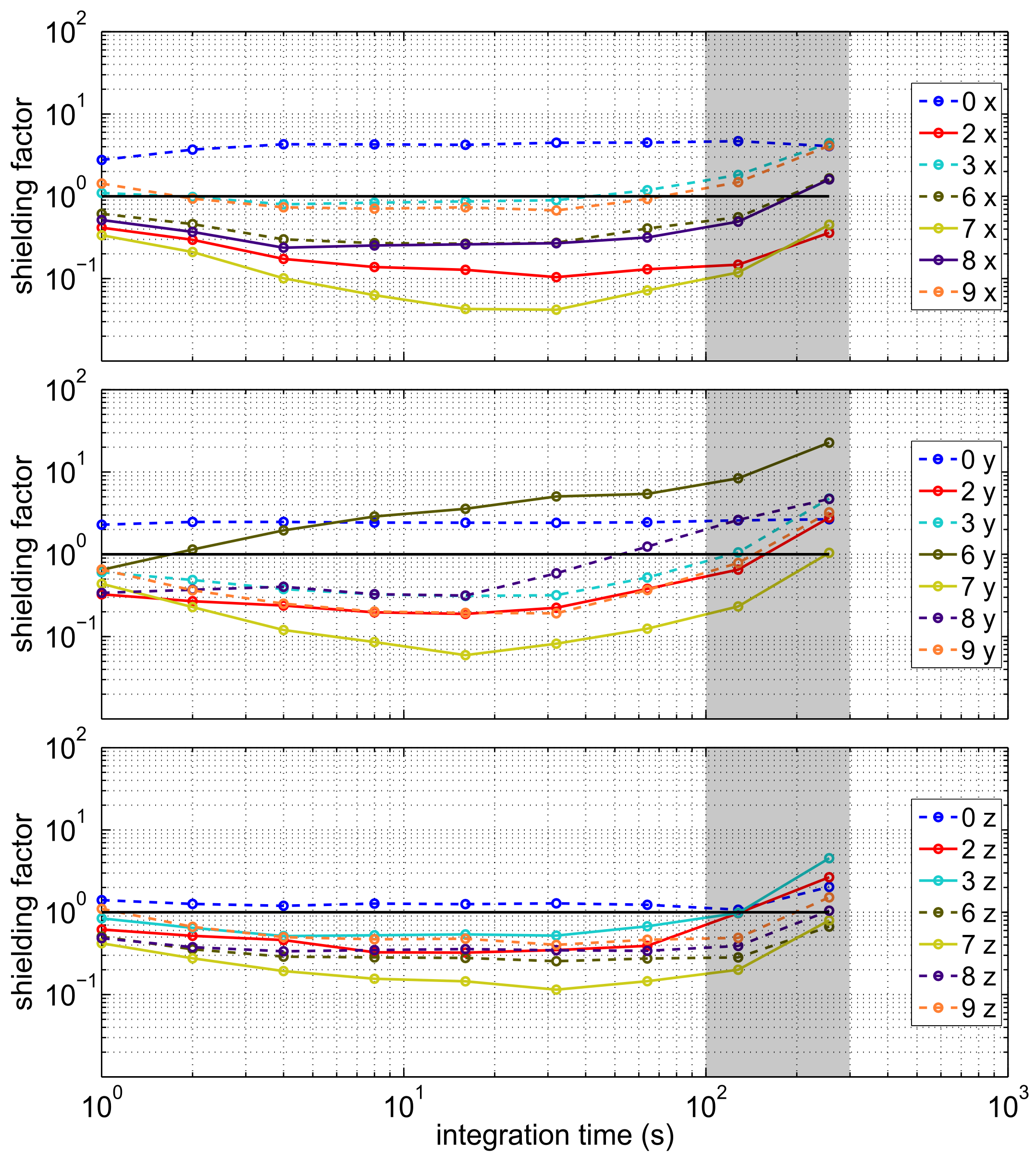}
\caption{Shielding factors from a measurement with twelve-sensor feedback 
with a non-regularized matrix.
Plot description as in Fig.\,\ref{fig:Sact-farfeedback}; 
feedback sensors plotted with dashed lines.}
\label{fig:Sact-unregpinv}
\end{figure} 

  \begin{figure} 
\centering
\includegraphics[angle=90,width=\OneColumnPictureWidth] {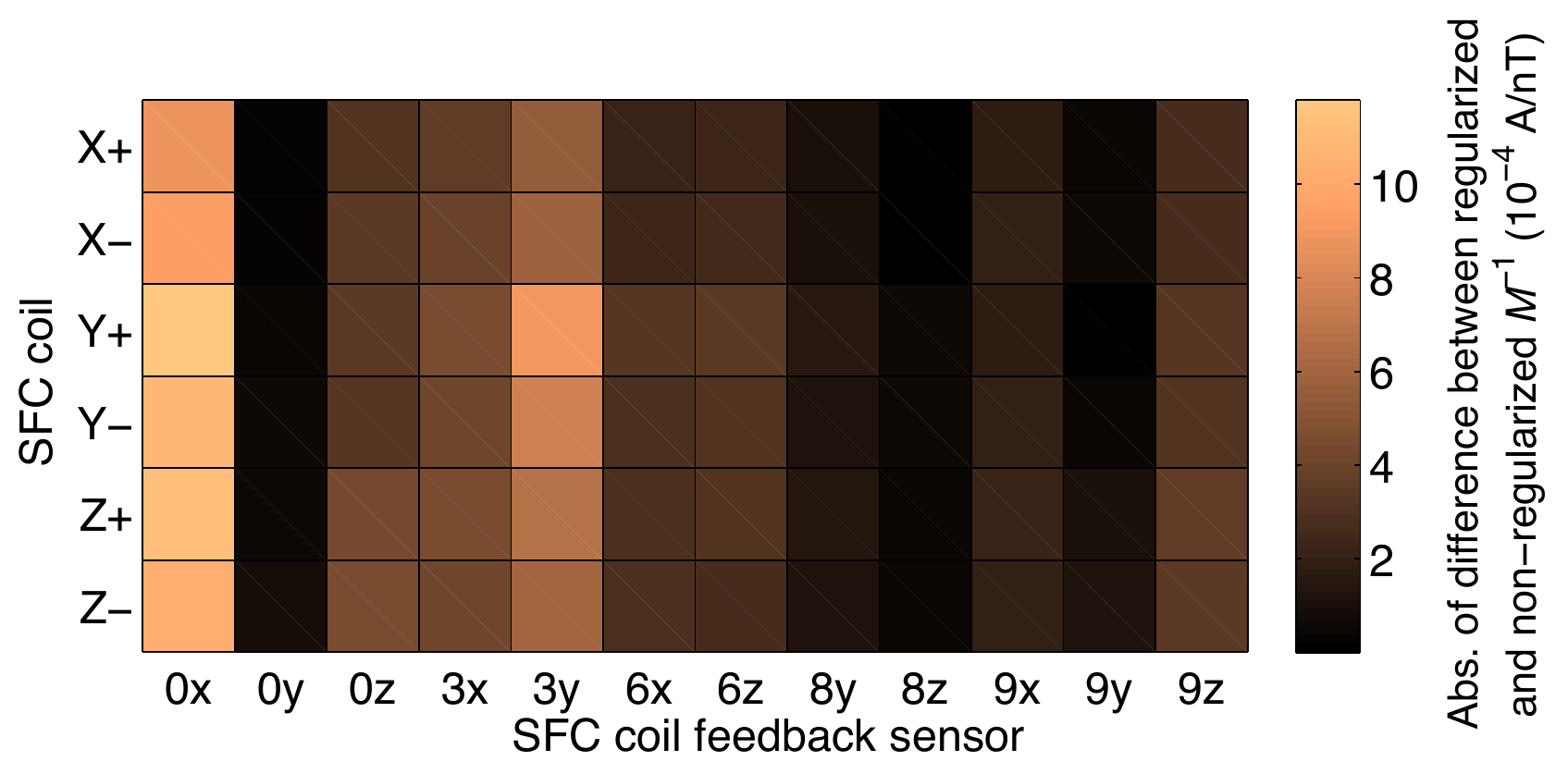}
\caption{Absolute value of the difference between
regularized $\hat{M}^{-1}_{jk,\mathrm{reg}}$  
and non-regularized $\hat{M}^{-1}_{jk}$ pseudoinverse 
of the proportionality matrix $M$
for the set of twelve feedback sensors used in the tests.
The value of the difference increases with increasing effect of regularization
on the specific matrix element.
Sensor $0x$ shows the largest effect.}
\label{fig:matrixdiff12}
\end{figure} 

\begin{figure}  
\centering
\includegraphics[width=\OneColumnPictureWidth]{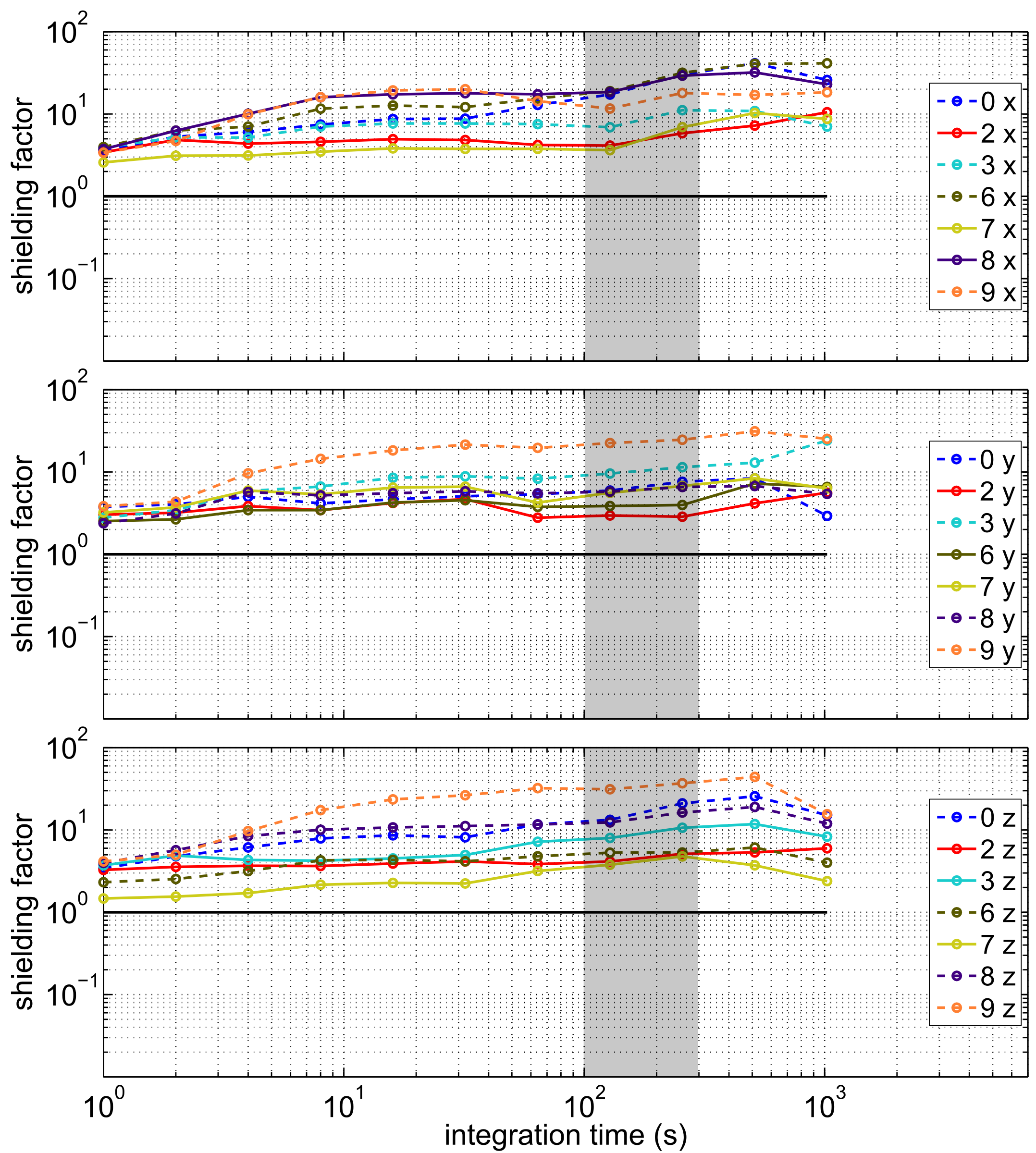}
\caption{Shielding factors from a measurement with 
twelve-sensor feedback including a regularized matrix.
These numbers can be compared to Fig.\,\ref{fig:Sact-unregpinv}, 
where the same feedback sensors were used without applying a 
regularization to the proportionality matrix. 
Plot description as in Fig.\,\ref{fig:Sact-farfeedback}; 
feedback sensors plotted with dashed lines.}
\label{fig:Sact-regpinv12}
\end{figure} 

The shielding factors $S_k$ measured with twelve feedback sensors, 
using a non-regularized pseudoinverse of the matrix of proportionality 
factors are shown in Fig.~\ref{fig:Sact-unregpinv}.
The feedback sensors 
were distributed over the entire control
volume with an equal number of x-, y-, and z-sensors.
The monitor sensors are depicted with solid lines.
Gains $\alpha^{\mathcal{P}}_j=\alpha^{\mathcal{I}}_j=0.5$ were used for all $j$.
The measured shielding factors were spread over a wide range, 
and most of them were below one.
Thus, the stability of the magnetic field was significantly decreased, 
caused by overcompensation of noise in sensor 0x, which showed a 
factor of four stability improvement.
In order to mitigate this overcompensation, a regularization with $r=3.0$ was applied.
The effect of the regularization on the matrix elements of the individual 
sensors is displayed in Fig.~\ref{fig:matrixdiff12} as a color-map of the 
absolute values of the differences $\left(\hat{M}^{-1}_{jk'}\right)^\mathrm{reg}-\hat{M}^{-1}_{jk'}$. 
Indeed, the matrix elements of sensor 0x are most affected by the regularization.

Dynamic feedback stabilization with the regularized matrix increased the shielding 
factors as shown in Fig.~\ref{fig:Sact-regpinv12}.
$\alpha^{\mathcal{P}}_j=1$ and $\alpha^{\mathcal{I}}_j=0.8$ were used for all $j$.
The smaller spread of the shielding factor values 
indicates that the stabilization effect by dynamic SFC was more homogeneous at different sensor positions.
The stability improved by factors of 4 to 30 at integration times greater than 10\,s 
at almost all sensor positions.
This demonstrates that using a regularized pseudoinverse matrix of proportionality
factors is an effective way to take into account the entanglement of all sensors and
coils and transfer the stabilization of the magnetic field at single feedback sensor
positions to a larger volume.


\subsection{SFC performance with 18 and 24 feedback sensors}
\label{sec:results18}

  \begin{figure}  
\centering
\includegraphics[width=\OneColumnPictureWidth] {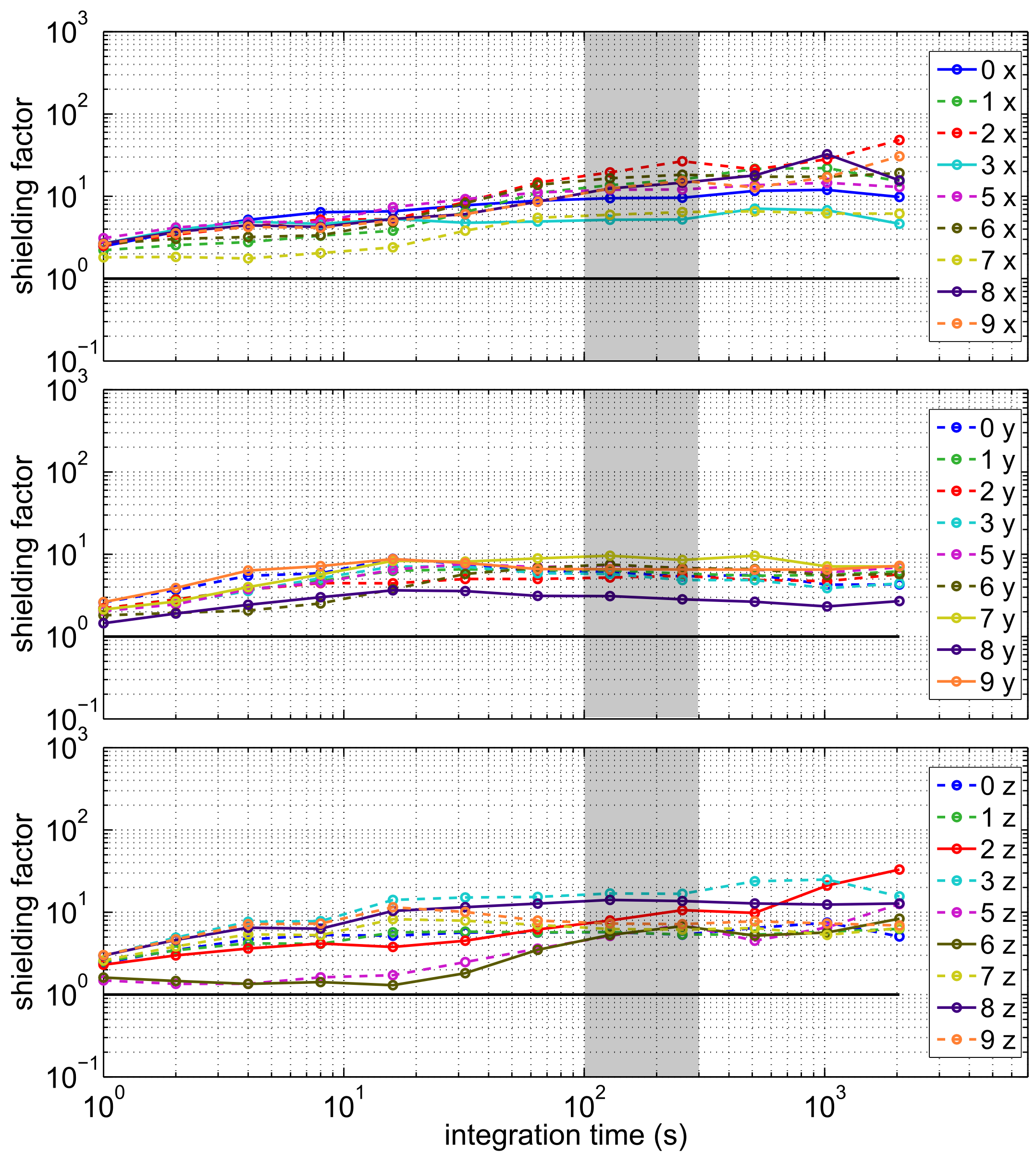}
\caption{Shielding factors from a typical measurement with 18-sensor 
feedback including a regularized matrix. 
Plot description as in Fig.\,\ref{fig:Sact-farfeedback}; feedback sensors are 
plotted with dashed lines.}
\label{fig:Sact-regpinv18}
\end{figure} 

We also investigated the influence of 18 and 24 
feedback sensors 
on the SFC stabilization performance.
Figure \ref{fig:Sact-regpinv18} shows the shielding factors achieved 
with 18 feedback sensors,
$r=3.4$,
$\alpha^{\mathcal{P}}_j=1$,
and $\alpha^{\mathcal{I}}_j=0.55$ for all $j$.
For feedback three sensors of
FG\,1 and FG\,5 and two sensors each of  
FG\,2, FG\,3, FG\,4, FG\,6, FG\,7 and FG\,9 were selected.
The achieved shielding factors
cover a range from 2 to 50 for $\tau>10\,$s,
comparable to the regularized case with 12 sensors.

\begin{figure}  
\centering
\includegraphics[width=\OneColumnPictureWidth]{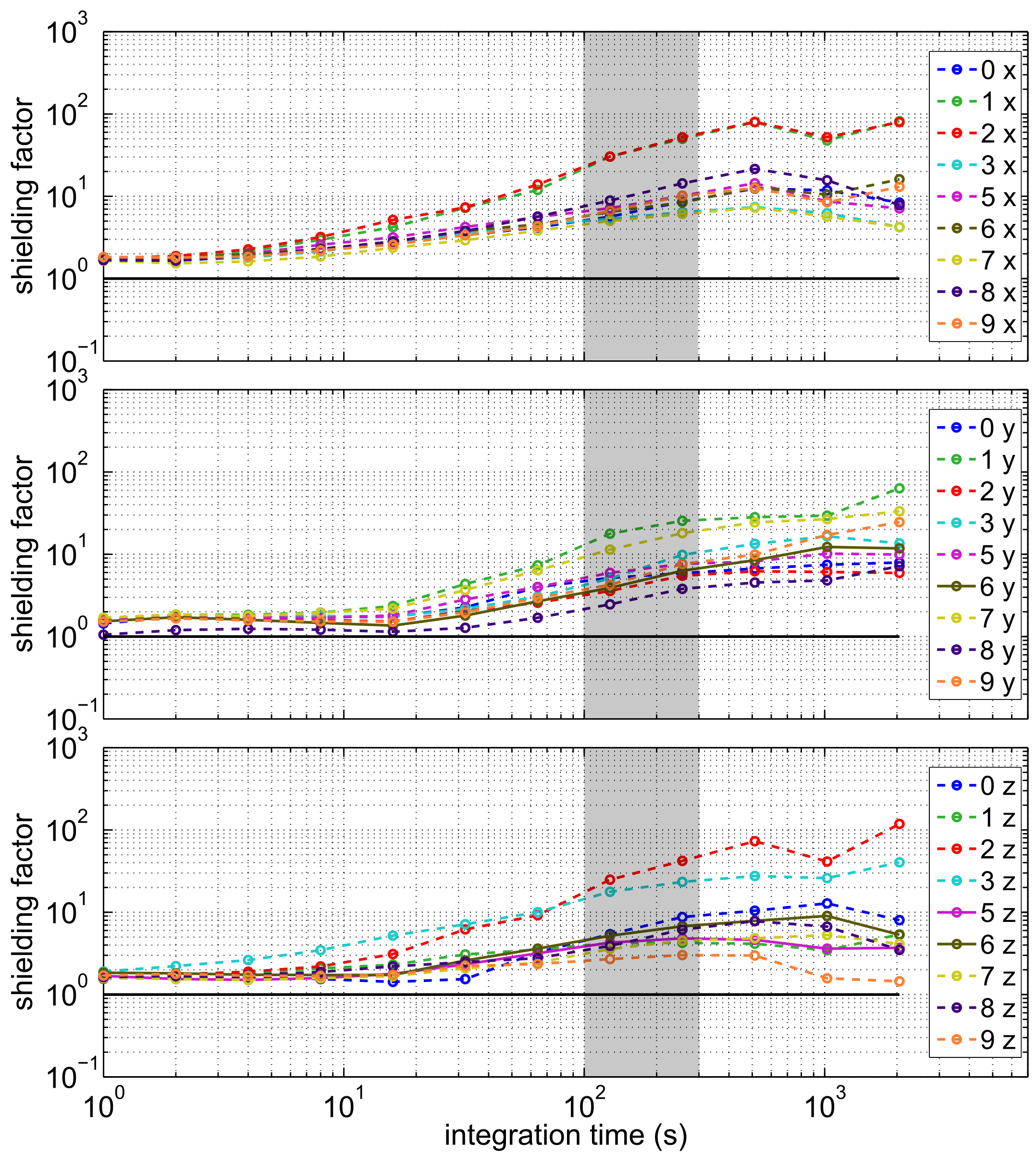}
\caption{Shielding factors from a typical measurement 
with 24-sensor feedback including a regularized matrix.
Plot description as in Fig.\,\ref{fig:Sact-farfeedback}; 
feedback sensors plotted with dashed lines.}
\label{fig:Sact-regpinv24}
\end{figure}

Figure~\ref{fig:Sact-regpinv24} shows the shielding factors achieved
with 24 feedback sensors, $r=3.4$, 
$\alpha^{\mathcal{P}}_j=0.9$, and $\alpha^{\mathcal{I}}_j=0.56$ for all $j$.
Only sensors 6y, 5z, and 6z were not used for the feedback.
The behavior is slightly different compared to the 18-sensor feedback.
The shielding factors are quite low for $\tau<10\,$s
which is probably caused by picking 
up noise of higher multipole order which cannot be compensated
by the present system.
Opening and closing of shutters or valves in the nEDM experiment
with operation times of a few seconds could be the source of this noise. 
For $\tau>20\,$s the shielding factors increased and reach a similar level 
as for the 12- and 18-sensor feedback.
The observed shielding factors agree with amplitude suppression 
of single-disturbance events, as e.g.~shown in Fig.\ref{fig:surrounding-field}.

\subsection{SFC performance with large field changes caused by remote sources}
\label{sec:sultanramp}

\begin{figure}  
\centering
\includegraphics[width=\OneColumnPictureWidth]{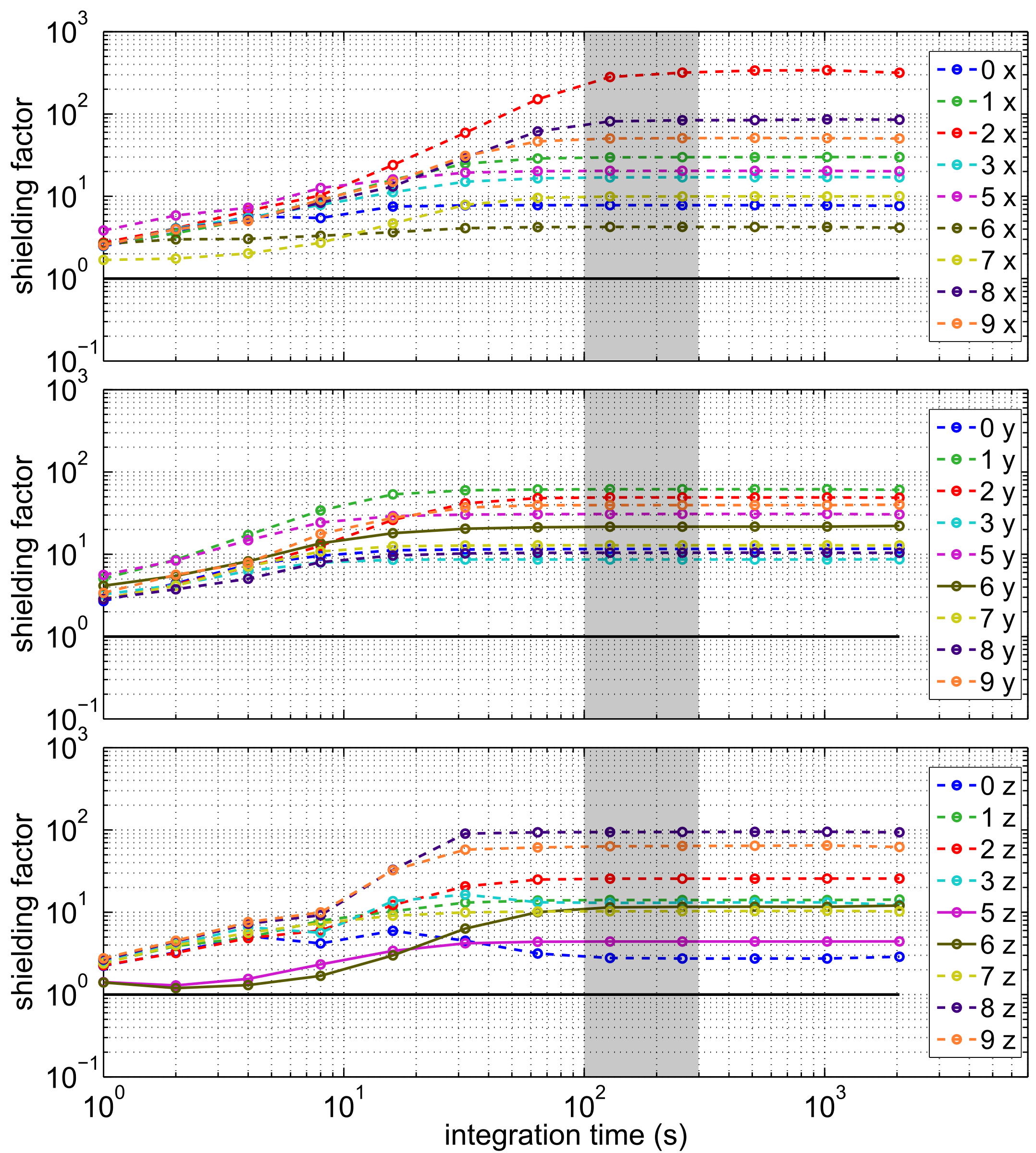}
\caption{Shielding factors from a typical compensation of the ramp of the 
SULTAN or EDIPO facility
with a 24-sensor feedback.
12- or 18-sensor feedback yielded comparable results.
Plot description as in Fig.\,\ref{fig:Sact-farfeedback}; 
feedback sensors plotted with dashed lines.}
\label{fig:Sact-Sultan}
\end{figure} 

The highest shielding factors 
observed so far were during the 
approximately hour-long 
magnetic field changes during ramping
of the neighboring superconducting magnets 
SULTAN or EDIPO.
These events cause a field change of up to 30\,$\mu$T 
at our experiment position in case of a full 12\,T ramp.
The two remote magnetic field sources 
differ in stray field magnitude, 
but have only small higher order multipoles at 
the position of our experiment.
Therefore, the SFC system 
can compensate for these perturbations very well, 
as shown in Fig.~\ref{fig:Sact-Sultan}.
Fig.~\ref{fig:ramp} shows the observed field values in a single sensor
during a SULTAN ramp outside and inside the Mu-metal shield and 
also the achieved compensation with the SFC system.
Nevertheless, the ramp can only be attenuated, 
the remaining field change, in spite of dynamic SFC, can be up to 2\,$\mu$T
at single sensor positions outside the shield.
A dedicated coil system which creates the particular 
compensation fields for SULTAN or EDIPO would be necessary in order to fully 
offset those ramps.
However, with an idealization procedure as described in \cite{Thiel2009},
the absolute value, as well as the three components of the magnetic field 
inside the Mu-metal shield, were reproduced to within a few hundreds of pT 
of the values before the ramp, as measured 
with scalar and vector magnetometers
inside the Mu-metal shield.
The observed shielding factors up to about 100 show the potential of the 
applied method for remote disturbances.


\section{Conclusions and outlook}
\label{sec:conclusion}

We have shown that the SFC reduces and stabilizes the magnetic field 
around and inside the Mu-metal shield of our apparatus.
This is important for conducting the nEDM measurements, specifically
in the time range from 100\,s to 300\,s.

When using a simple feedback algorithm without implementation of a matrix of proportionality factors, 
high shielding factors were achieved only at the locations of the chosen feedback sensors.
The obtained shielding factors in the control volume ranged from 2 to 5.
At magnetically quiet times, such a simple feedback type even decreased the field stability slightly.

The shielding factors were increased to values of 3 to 50 by including a 
regularized pseudoinverse matrix of proportionality factors.
No significant difference was observed in the quality of the magnitude of the shielding
factors at stabilized and non-stabilized sensor positions.
Furthermore, comparable results were achieved at magnetically noisy periods and at quiet times.

We have shown that in our setup the shielding factors do not improve 
when the number of feedback sensors is 
increased from 12 to 18 or to 24.
On the contrary, the shielding factors for short integration 
times ($\tau<10\,$s) decrease with increasing number of feedback sensors,
which may  
pick up very localized higher-order multipole magnetic noise.

In the case of remote magnetic disturbances containing no, 
or only small higher-order, 
multipole contributions, shielding factors of up to 100 were achieved.

The performance of the SFC system could be extended to compensate for
higher multipole field perturbations by increasing the number of coils in the system.
R\&D for systems with a larger number of coils and field sensors are being pursued,
together with further refinement of the SFC feedback model.


\section{Acknowledgements}

\label{acknow}

The Paul Scherrer Institute and ETH Z\"urich gratefully acknowledge
the support of the 
Swiss National Science Foundation under Grant Nos.
200021\_126562, 200020\_144473, 200021\_138211,
200020\_149211
and CRSII2\_144257.
University of Fribourg acknowledges financial support 
by Grant 200020\_140421 of the Swiss National Science Foundation.
KU Leuven acknowledges the support of the Fund for Scientific 
Research Flanders (FWO) and Project GOA/2010/10 of the KU Leuven. 
LPC Caen and LPSC Grenoble acknowledge Grant No. ANR-09-BLAN-0046.
Jagellonian University Cracow acknowledges support by the Foundation for Polish Science - MPD 
program, co-financed by the European Union within the European Regional Development Fund,
and by the National Science Centre, Poland, under Grant No. 2012/04/M/ST2/00556.
The contribution of TU M\"unchen in the beginning of the project is acknowledged.


\hbadness=10000  

\bibliography{papersSFC}

\end{document}